\documentclass[aps,prd,twocolumn,floatfix,
10pt,
amssymb,amsfont,amsmath,
float,nofootinbib,longbibliography]{revtex4-1}

\usepackage[utf8]{inputenc}
\usepackage{lmodern}
\usepackage{verbatim}
\usepackage[usenames,dvipsnames]{xcolor}
\usepackage{graphicx}
\usepackage[pdftex, plainpages=false]{hyperref}
\usepackage{mathtools}
\usepackage{stmaryrd, ifthen}
\usepackage{array}
\usepackage{tabularx,float}
\usepackage[export]{adjustbox}
\usepackage{mathrsfs, bbm, yfonts, wasysym, dsfont}
\usepackage{enumitem}
\usepackage{calc}
\usepackage{bm}
\usepackage{etoolbox}
\mathtoolsset{showonlyrefs=false}
\usepackage{footnote}
\usepackage{titlesec}
\usepackage{dcolumn}
\usepackage{microtype}
\usepackage{booktabs}
\usepackage{ upgreek }

\usepackage{tikz}
\usetikzlibrary{shapes.geometric}
\usetikzlibrary{arrows,matrix,calc,scopes,decorations.markings}
\usetikzlibrary{arrows.meta}
\usetikzlibrary{intersections, patterns}
\usetikzlibrary{decorations.pathreplacing,angles,quotes}
\usetikzlibrary{external}
\tikzsetexternalprefix{fig/}

\titleformat{name=\section}
{\normalfont}
{\centering {\textsc{Sec.} \thesection \; \raisebox{1pt}{\textbar} \;}}
{0pt}
{\normalsize\bfseries\centering}

\titleformat{name=\subsection}
{\small}
{\centering {\thesection .\thesubsection . \;}}
{0pt}
{\small\bfseries\centering}

\makeatletter
\def\l@subsubsection#1#2{}
\makeatother

\tikzset{
	every picture/.append style={
		line join=round,
		line cap=round,
		line width = 0.6pt
	}
}

\tikzset{->1-/.style={decoration={
			markings,
			mark=at position #1 with {\arrow{Triangle[fill=black, width=3.5pt, length=3.5pt]}}},postaction={decorate}}}

\tikzstyle{op9}=[opacity=0.6]
\tikzstyle{op8}=[opacity=0.6]
\tikzstyle{op7}=[opacity=0.6]
\tikzstyle{op6}=[opacity=0.6]
\tikzstyle{op5}=[opacity=0.5]
\tikzstyle{op4}=[opacity=0.4]
\tikzstyle{op3}=[opacity=0.3]
\tikzstyle{op2}=[opacity=0.2]
\tikzstyle{op1}=[opacity=0.1]

\tikzstyle{sha}=[shorten >= 2.5pt]
\tikzstyle{shb}=[shorten <= 2.5pt]
\tikzstyle{shab}=[shorten >= 2.5pt, shorten <= 2.5pt]

\tikzstyle{tenBG}=[fill = white, fill opacity=1]
\tikzstyle{ten0}=[draw=black, fill = BlueViolet, fill opacity=0.35]
\tikzstyle{ten0b}=[draw=black, fill = BlueViolet, fill opacity=0.6]
\tikzstyle{ten0c}=[draw=black, fill = BlueViolet, fill opacity=0.1]
\tikzstyle{ten0d}=[draw=black, fill = BlueViolet, fill opacity=0.2]
\tikzstyle{ten1}=[draw=black, fill = BrickRed, fill opacity=0.5]
\tikzstyle{ten1b}=[draw=black, fill = BrickRed, fill opacity=0.3]
\tikzstyle{ten2}=[draw=black, fill = NavyBlue, fill opacity=0.35]
\tikzstyle{ten3}=[draw=black, fill = LimeGreen, fill opacity=0.5]
\tikzstyle{ten4}=[draw=black, fill = BurntOrange, fill opacity=0.5]
\tikzstyle{ten5}=[draw=black, fill = Thistle, fill opacity=0.5]
\tikzstyle{specCirc}=[circle, fill=BlueViolet, opacity=0.5, inner sep=1pt]

\newcommand{\interLL}[7]{
	\path[name path=P#1#2] (#1) -- (#2);
	\path[name path=P#3#4] (#3) -- (#4);
	\path[name path=P#5#6] (#5) -- (#6);	
	\path [name intersections={of=P#1#2 and P#3#4,by={sp1}}];
	\path [name intersections={of=P#1#2 and P#5#6,by={sp2}}];	
	\draw[sha, #7] (#1) -- (sp1);
	\draw[shab, #7] (sp1) -- (sp2);
	\draw[shb, #7] (sp2) -- (#2);
}

\newcommand{\ten}[2]{
    \centerarc[tenBG](#1)(0:360:0.16)
    \centerarc[#2](#1)(0:360:0.16)
}

\newcommand{\CDL}[2]{
    \centerarc[double, double distance = 0.5pt,line width = 0.6pt, opacity=0.25, color=BrickRed](#1)(0:360:#2)
}

\newcommand{\brokenCDL}[2]{
    \centerarc[double, double distance = 0.5pt,line width = 0.6pt, opacity=0.25, color=BrickRed](#1)(15:75:#2)
    \centerarc[double, double distance = 0.5pt,line width = 0.6pt, opacity=0.25, color=BrickRed](#1)(105:165:#2)
    \centerarc[double, double distance = 0.5pt,line width = 0.6pt, opacity=0.25, color=BrickRed](#1)(195:255:#2)
    \centerarc[double, double distance = 0.5pt,line width = 0.6pt, opacity=0.25, color=BrickRed](#1)(285:345:#2)
}

\newcommand{\deltaTen}[1]{
    \centerarc[draw=black, fill=black](#1)(0:360:0.07)
}

\newcommand{\tenBig}[2]{
    \centerarc[tenBG](#1)(0:360:0.22)
    \centerarc[#2](#1)(0:360:0.22)
}

\newcommand{\tenSmall}[4]{
    \centerarc[tenBG](#1,#2)(0:360:0.14)
    \centerarc[#3](#1,#2)(0:360:0.14)
    \ifthenelse{#4 = 1}{
        \draw[line width =0.4pt] ({#1-0.07},{#2-0.07}) rectangle ({#1+0.07},{#2+0.07});
    }{}
    \ifthenelse{#4 = 2}{
        \draw[line width =0.4pt] ({#1-sqrt(0.14^2/3)},{#2-0.07+0.01}) -- ({#1+sqrt(0.14^2/3)},{#2-0.07+0.01}) -- (#1,{#2+0.07+0.01}) -- ({#1-sqrt(0.14^2/3)},{#2-0.07+0.01});
    }{}
    \ifthenelse{#4 = 3}{
        \draw[line width =0.4pt] ({#1-0.07},{#2-0.07}) -- ({#1+0.07},{#2+0.07});
        \draw[line width =0.4pt] ({#1-0.07},{#2+0.07}) -- ({#1+0.07},{#2-0.07});
    }{}
    \ifthenelse{#4 = 4}{
        \centerarc[line width =0.4pt](#1,#2)(0:360:0.07)
    }{}
}

\newcommand{\halfTenSmallL}[4]{
    \centerarc[tenBG](#1,#2)(90:270:0.14)
    \centerarc[#3](#1,#2)(90:270:0.14)
    \begin{scope}
        \clip (#1,{#2+0.14}) rectangle ({#1-0.14},{#2-0.14});
        \ifthenelse{#4 = 2}{
            \draw[line width =0.4pt] ({#1-sqrt(0.14^2/3)},{#2-0.07+0.01}) -- ({#1+sqrt(0.14^2/3)},{#2-0.07+0.01}) -- (#1,{#2+0.07+0.01}) -- ({#1-sqrt(0.14^2/3)},{#2-0.07+0.01});
        }{}
        \ifthenelse{#4 = 3}{
            \draw[line width =0.4pt] ({#1-0.07},{#2-0.07}) -- ({#1+0.07},{#2+0.07});
            \draw[line width =0.4pt] ({#1-0.07},{#2+0.07}) -- ({#1+0.07},{#2-0.07});
        }{}
    \end{scope}
    \draw[] (#1,{#2+0.14}) -- (#1,{#2-0.14});
}

\newcommand{\halfTenSmallR}[4]{
    \centerarc[tenBG](#1,#2)(-90:90:0.14)
    \centerarc[#3](#1,#2)(-90:90:0.14)
    \begin{scope}
        \clip (#1,{#2+0.14}) rectangle ({#1+0.14},{#2-0.14});
        \ifthenelse{#4 = 2}{
            \draw[line width =0.4pt] ({#1-sqrt(0.14^2/3)},{#2-0.07+0.01}) -- ({#1+sqrt(0.14^2/3)},{#2-0.07+0.01}) -- (#1,{#2+0.07+0.01}) -- ({#1-sqrt(0.14^2/3)},{#2-0.07+0.01});
        }{}
        \ifthenelse{#4 = 3}{
            \draw[line width =0.4pt] ({#1-0.07},{#2-0.07}) -- ({#1+0.07},{#2+0.07});
            \draw[line width =0.4pt] ({#1-0.07},{#2+0.07}) -- ({#1+0.07},{#2-0.07});
        }{}
    \end{scope}
    \draw[] (#1,{#2+0.14}) -- (#1,{#2-0.14});
}

\newcommand{\halfTenSmallB}[4]{
    \centerarc[tenBG](#1,#2)(0:-180:0.14)
    \centerarc[#3](#1,#2)(0:-180:0.14)
    \begin{scope}
        \clip ({#1+0.14},#2) rectangle ({#1-0.14},{#2-0.14});
        \ifthenelse{#4 = 1}{
            \draw[line width =0.4pt] ({#1-0.07},{#2-0.07}) rectangle ({#1+0.07},{#2+0.07});
        }{}
        \ifthenelse{#4 = 4}{
            \centerarc[line width =0.4pt](#1,#2)(0:360:0.07)
        }{}
    \end{scope}
    \draw[] ({#1-0.14},#2) -- ({#1+0.14},#2);
}

\newcommand{\halfTenSmallT}[4]{
    \centerarc[tenBG](#1,#2)(0:180:0.14)
    \centerarc[#3](#1,#2)(0:180:0.14)
    \begin{scope}
        \clip ({#1+0.14},#2) rectangle ({#1-0.14},{#2+0.14});
        \ifthenelse{#4 = 1}{
            \draw[line width =0.4pt] ({#1-0.07},{#2-0.07}) rectangle ({#1+0.07},{#2+0.07});
        }{}
        \ifthenelse{#4 = 4}{
            \centerarc[line width =0.4pt](#1,#2)(0:360:0.07)
        }{}
    \end{scope}
    \draw[] ({#1-0.14},#2) -- ({#1+0.14},#2);
}

\newcommand{\halfTenL}[2]{
    \centerarc[tenBG](#1)(90:270:0.16)
    \centerarc[#2](#1)(90:270:0.16)
    \draw[] (#1) --++(0,0.16);
    \draw[] (#1) --++(0,-0.16); 
}

\newcommand{\halfTenR}[2]{
    \centerarc[tenBG](#1)(-90:90:0.16)
    \centerarc[#2](#1)(-90:90:0.16)
    \draw[] (#1) --++(0,0.16);
    \draw[] (#1) --++(0,-0.16); 
}

\newcommand{\halfTenTL}[2]{
    \centerarc[tenBG](#1)(-135:45:0.16)
    \centerarc[#2](#1)(-135:45:0.16)
    \draw[] (#1) --++({-sqrt(0.5*0.16^2)},{-sqrt(0.5*0.16^2)});
    \draw[] (#1) --++({sqrt(0.5*0.16^2)},{sqrt(0.5*0.16^2)}); 
}

\newcommand{\halfTenTR}[2]{
    \centerarc[tenBG](#1)(-45:-225:0.16)
    \centerarc[#2](#1)(-45:-225:0.16)
    \draw[] (#1) --++({sqrt(0.5*0.16^2)},{-sqrt(0.5*0.16^2)});
    \draw[] (#1) --++({-sqrt(0.5*0.16^2)},{sqrt(0.5*0.16^2)}); 
}

\newcommand{\halfTenBL}[2]{
    \centerarc[tenBG](#1)(-45:135:0.16)
    \centerarc[#2](#1)(-45:135:0.16)
    \draw[] (#1) --++({sqrt(0.5*0.16^2)},{-sqrt(0.5*0.16^2)});
    \draw[] (#1) --++({-sqrt(0.5*0.16^2)},{sqrt(0.5*0.16^2)}); 
}

\newcommand{\halfTenBR}[2]{
    \centerarc[tenBG](#1)(45:225:0.16)
    \centerarc[#2](#1)(45:225:0.16)
    \draw[] (#1) --++({sqrt(0.5*0.16^2)},{sqrt(0.5*0.16^2)});
    \draw[] (#1) --++({-sqrt(0.5*0.16^2)},{-sqrt(0.5*0.16^2)}); 
}

\newcommand{\TRGUnitCell}[2]{
    \draw[] ({#1-0.5},{#2-0.5}) -- ({#1-0.35},{#2-0.35});
    \draw[] ({#1-0.5},{#2+0.5}) -- ({#1-0.35},{#2+0.35});
    \draw[] ({#1+0.5},{#2-0.5}) -- ({#1+0.35},{#2-0.35});
    \draw[] ({#1+0.5},{#2+0.5}) -- ({#1+0.35},{#2+0.35});
    \draw[] ({#1-0.35},{#2-0.35}) rectangle ({#1+0.35},{#2+0.35});
    \halfTenTL{{#1-0.35},{#2+0.35}}{ten0};
    \halfTenTR{{#1+0.35},{#2+0.35}}{ten0};
    \halfTenBL{{#1-0.35},{#2-0.35}}{ten0};
    \halfTenBR{{#1+0.35},{#2-0.35}}{ten0};
}

\newcommand{\TNRUnitCell}[3]{
    \ifthenelse{#3 = 1}{
        \tenSmall{#1-0.5}{#2}{ten4}{1};
        \tenSmall{#1+0.5}{#2}{ten4}{4};
        \tenSmall{#1}{#2-0.5}{ten4}{3};
        \tenSmall{#1}{#2+0.5}{ten4}{2};
    }{}
    \ifthenelse{#3 = 2}{
        \halfTenSmallT{#1-0.5}{#2+0.08}{ten4}{1};
        \halfTenSmallB{#1-0.5}{#2-0.08}{ten4}{1};
        \halfTenSmallT{#1+0.5}{#2+0.08}{ten4}{4};
        \halfTenSmallB{#1+0.5}{#2-0.08}{ten4}{4};
        \halfTenSmallL{#1-0.08}{#2-0.5}{ten4}{3};
        \halfTenSmallR{#1+0.08}{#2-0.5}{ten4}{3};
        \halfTenSmallL{#1-0.08}{#2+0.5}{ten4}{2};
        \halfTenSmallR{#1+0.08}{#2+0.5}{ten4}{2};
    }{}
}

\newcommand{\OGUnitCell}[2]{
    \halfTenSmallT{#1-0.5}{#2+0.12}{ten0c}{0};
    \halfTenSmallB{#1-0.5}{#2-0.12}{ten0c}{0};
    \halfTenSmallT{#1+0.5}{#2+0.12}{ten0c}{0};
    \halfTenSmallB{#1+0.5}{#2-0.12}{ten0c}{0};
    \halfTenSmallL{#1-0.12}{#2-0.5}{ten0c}{0};
    \halfTenSmallR{#1+0.12}{#2-0.5}{ten0c}{0};
    \halfTenSmallL{#1-0.12}{#2+0.5}{ten0c}{0};
    \halfTenSmallR{#1+0.12}{#2+0.5}{ten0c}{0};
}

\newcommand{\TRGUnitCellBi}[4]{
    \draw[] ({#1-0.5},{#2-0.5}) -- ({#1-0.35},{#2-0.35});
    \draw[] ({#1-0.5},{#2+0.5}) -- ({#1-0.35},{#2+0.35});
    \draw[] ({#1+0.5},{#2-0.5}) -- ({#1+0.35},{#2-0.35});
    \draw[] ({#1+0.5},{#2+0.5}) -- ({#1+0.35},{#2+0.35});
    \draw[] ({#1-0.35},{#2-0.35}) rectangle ({#1+0.35},{#2+0.35});
    \halfTenTL{{#1-0.35},{#2+0.35}}{#3};
    \halfTenTR{{#1+0.35},{#2+0.35}}{#4};
    \halfTenBL{{#1-0.35},{#2-0.35}}{#4};
    \halfTenBR{{#1+0.35},{#2-0.35}}{#3};
}

\newcommand{\lattice}[4]{
    \foreach \a in {#1,...,#3}
        \draw[] (\a,#2) -- (\a,#4);
    \foreach \b in {#2,...,#4}
        \draw[] (#1,\b) -- (#3,\b);    
}

\newcommand{\network}[5]{
    \foreach \x in {#1,...,#3}
	    \foreach \y in {#2,...,#4}
	        \ten{\x,\y}{#5};    
}

\newcommand{\networkBi}[2]{
    \foreach \x in {0,2}
        \foreach \y in {0,2}
            \ten{\x,\y}{#1};
    \foreach \x in {1,3}
        \foreach \y in {1,3}
            \ten{\x,\y}{#1};
    \foreach \x in {1,3}
        \foreach \y in {0,2}
            \ten{\x,\y}{#2};
    \foreach \x in {0,2}
        \foreach \y in {1,3}
            \ten{\x,\y}{#2};          
}

\def\centerarc[#1](#2)(#3:#4:#5)
{ \draw[#1] ($(#2)+({#5*cos(#3)},{#5*sin(#3)})$) arc (#3:#4:#5); }

\newcommand{\tadpole}[0]{
	\begin{tikzpicture}[scale=0.85,baseline=-0.2em]
        \draw[] (0,0) -- (2,0);
        \draw[] (1,0) to[out=10, in=0, looseness=1.5] (1,1) to[out=-180, in=170, looseness=1.5] (1,0);
        \node[circle, fill=black, inner sep=1.5pt] at (1,0) {};
	\end{tikzpicture}
}

\newcommand{\deltaInit}[0]{
    \begin{tikzpicture}[scale=0.85,baseline={([yshift=-.5ex]current bounding box.center)}]
        \draw[] (-0.5,0) -- (0.5,0);
        \draw[] (0,0.5) -- (0,-0.5);
        \deltaTen{0,0};
        \node[left] at (-0.5,0) {${\phi_\ell}$};
        \node[above] at (0,0.5) {${\phi_u}$};
        \node[right] at (0.5,0) {${\phi_r}$};
        \node[below] at (0,-0.5) {${\phi_d}$};
    \end{tikzpicture}
}

\newcommand{\matrixInit}[1]{
    \begin{tikzpicture}[scale=0.85,baseline=-0.2em]
        \ifthenelse{#1 = 1}{
            \draw[] (0,0) -- (1,0);
            \tenSmall{0.5}{0}{ten0c}{0};
            \node[above, yshift=3pt] at (0.5,0) {${F}$};
            \node[left] at (0,0) {${\phi_\ell}$};
            \node[right] at (1,0) {${\phi_r}$};
        }{}
        \ifthenelse{#1 = 2}{
            \draw[] (0,0) -- (1,0);
            \halfTenSmallL{0.5-0.12}{0}{ten0c}{0};
            \halfTenSmallR{0.5+0.12}{0}{ten0c}{0};
            \node[above, yshift=3pt] at (0.5,0) {${\sss n}$};
            \node[left] at (0,0) {${\phi_\ell}$};
            \node[right] at (1,0) {${\phi_r}$};
        }{}
        \ifthenelse{#1 = 4}{
            \draw[] (0,0) -- (1,0);
            \halfTenSmallL{0.5-0.12}{0}{ten0c}{0};
            \halfTenSmallR{0.5+0.12}{0}{ten0c}{0};
            \node[left] at (0,0) {${\phi_\ell}$};
            \node[right] at (1,0) {${\phi_r}$};
        }{}
        \ifthenelse{#1 = 3}{
            \draw[] (0,0) -- (0.5,0); 
            \halfTenSmallL{0.5-0.12}{0}{ten0c}{0};
            \node[left] at (0,0) {${\phi}$};
            \node[right] at (0.5,0) {${\sss n}$};
        }{}
    \end{tikzpicture}
}

\newcommand{\CDLTensor}[0]{
    \begin{tikzpicture}[scale=0.85,baseline=-0.2em]
        \draw[double, double distance = 0.5pt,line width = 0.6pt, opacity=0.25, color=BrickRed] (-0.5,0.07) to[out=0, in=-90, looseness=1.5] (-0.07,0.5);
        \draw[double, double distance = 0.5pt,line width = 0.6pt, opacity=0.25, color=BrickRed] (-0.5,-0.07) to[out=0, in=90, looseness=1.5] (-0.07,-0.5);
        \draw[double, double distance = 0.5pt,line width = 0.6pt, opacity=0.25, color=BrickRed] (0.5,0.07) to[out=180, in=-90, looseness=1.5] (0.07,0.5);
        \draw[double, double distance = 0.5pt,line width = 0.6pt, opacity=0.25, color=BrickRed] (0.5,-0.07) to[out=180, in=90, looseness=1.5] (0.07,-0.5);
    \end{tikzpicture}
}

\newcommand{\networkOG}[1]{
	\begin{tikzpicture}[scale=0.85,baseline={([yshift=-.5ex]current bounding box.center)}]
	\ifthenelse{#1 = 1}{
    	\begin{scope}
        	\clip (-0.5,-0.5) rectangle (3.5,3.5);
            \lattice{-1}{-1}{4}{4}	    
            \foreach \x in {0,...,3}
	            \foreach \y in {0,...,3}
	                \deltaTen{\x,\y};  
            \foreach \x in {0,...,3}
                \foreach \y in {-0.5,0.5,1.5,2.5,3.5}
                    \tenSmall{\x}{\y}{ten0c}{0}; 
            \foreach \x in {-0.5,0.5,1.5,2.5,3.5}
                \foreach \y in {0,...,3}
                    \tenSmall{\x}{\y}{ten0c}{0};           
    	\end{scope}
	}{}
	\ifthenelse{#1 = 2}{
    	\begin{scope}
        	\clip (-0.5,-0.5) rectangle (3.5,3.5);
            \lattice{-1}{-1}{4}{4}	    
            \foreach \x in {0,...,3}
	            \foreach \y in {0,...,3}
	                \deltaTen{\x,\y};  
            \foreach \x in {0.5,2.5}
                \foreach \y in {0.5,2.5}
                    \OGUnitCell{\x}{\y}; 
            \foreach \x in {-0.5,1.5,3.5}
                \foreach \y in {-0.5,1.5,3.5}
                    \OGUnitCell{\x}{\y};           
    	\end{scope}
	}{}
	\end{tikzpicture}
}

\newcommand{\elementTensor}[1]{
    \begin{tikzpicture}[scale=0.85,baseline={([yshift=-.5ex]current bounding box.center)}]
        \draw[] (-0.5,0) -- (0.5,0);
        \draw[] (0,0.5) -- (0,-0.5);
        \ifthenelse{#1 = 1}{
            \ten{0,0}{ten0};
        }{}
        \ifthenelse{#1 = 2}{
            \deltaTen{0,0};
            \halfTenSmallB{0}{0.5-0.12}{ten0c}{0};
            \halfTenSmallT{0}{-0.5+0.12}{ten0c}{0};
            \halfTenSmallR{-0.5+0.12}{0}{ten0c}{0};
            \halfTenSmallL{0.5-0.12}{0}{ten0c}{0};
        }{}
        \node[left, xshift=1pt] at (-0.5,0) {${\sss a}$};
        \node[above, yshift=-1pt] at (0,0.5) {${\sss b}$};
        \node[right, xshift=-1pt] at (0.5,0) {${\sss c}$};
        \node[below, yshift=1pt] at (0,-0.5) {${\sss d}$};
    \end{tikzpicture}
}

\newcommand{\SVD}[1]{
	\begin{tikzpicture}[scale=0.85,baseline={([yshift=-.5ex]current bounding box.center)}]
	    \ifthenelse{#1 = 1}{
            \draw[] (0,0) -- (-0.5,0.5);
            \draw[] (0,0) -- (-0.5,-0.5);
            \draw[] (0,0) -- (0.5,0.5);
            \draw[] (0,0) -- (0.5,-0.5);
            \ten{0,0}{ten1};
            \draw[loosely dotted, shorten <= 4pt, shorten >= 4pt] (-0.4,0.4) to[out=-110, in=110, looseness=1] (-0.4,-0.4);
            \draw[loosely dotted, shorten <= 4pt, shorten >= 4pt] (0.4,0.4) to[out=-70, in=70, looseness=1] (0.4,-0.4);
            \node[above, yshift=3pt] at (0,0) {$A$};
        }{}
        \ifthenelse{#1 = 2}{
            \draw[] (0,0) -- (-0.5,0.5);
            \draw[] (0,0) -- (-0.5,-0.5);
            \draw[] (1.2,0) -- (1.7,0.5);
            \draw[] (1.2,0) -- (1.7,-0.5);
            \draw[] (0,0) -- (1.2,0);
            \ten{0,0}{ten1b};
            \ten{1.2,0}{ten1b};
            \ten{0.6,0}{ten3};
            \draw[loosely dotted, shorten <= 4pt, shorten >= 4pt] (-0.4,0.4) to[out=-110, in=110, looseness=1] (-0.4,-0.4);
            \draw[loosely dotted, shorten <= 4pt, shorten >= 4pt] (1.6,0.4) to[out=-70, in=70, looseness=1] (1.6,-0.4);
            \node[above, yshift=3pt] at (0,0) {$U$};
            \node[below, yshift=-3pt] at (0.6,0) {$S$};
            \node[above, yshift=3pt] at (1.2,0) {$V^\dagger$};
        }{}
        \ifthenelse{#1 = 3}{
            \draw[] (0,0) -- (-0.5,0.5);
            \draw[] (0,0) -- (-0.5,-0.5);
            \draw[] (1.2,0) -- (1.7,0.5);
            \draw[] (1.2,0) -- (1.7,-0.5);
            \draw[] (0,0) -- (1.2,0);
            \ten{0,0}{ten1b};
            \ten{1.2,0}{ten1b};
            \halfTenL{0.5,0}{ten3};
            \halfTenR{0.7,0}{ten3};
            \draw[loosely dotted, shorten <= 4pt, shorten >= 4pt] (-0.4,0.4) to[out=-110, in=110, looseness=1] (-0.4,-0.4);
            \draw[loosely dotted, shorten <= 4pt, shorten >= 4pt] (1.6,0.4) to[out=-70, in=70, looseness=1] (1.6,-0.4);
            \node[above, yshift=3pt] at (0,0) {$U$};
            \node[above, yshift=3pt] at (1.2,0) {$V^\dagger$};
            \node[below, yshift=-3pt, xshift=-5pt] at (0.5,0) {$S^\frac{1}{2}$};
            \node[below, yshift=-3pt, xshift=5pt] at (0.7,0) {$S^\frac{1}{2}$};
        }{}
        \ifthenelse{#1 = 4}{
            \draw[] (0,0) -- (-0.5,0.5);
            \draw[] (0,0) -- (-0.5,-0.5);
            \draw[] (0,0) -- (0.6,0);
            \ten{0,0}{ten1b};
            \halfTenL{0.5,0}{ten3};
            \draw[loosely dotted, shorten <= 4pt, shorten >= 4pt] (-0.4,0.4) to[out=-110, in=110, looseness=1] (-0.4,-0.4);
        }{}  
        \ifthenelse{#1 = 5}{
            \draw[] (0,0) -- (-0.5,0.5);
            \draw[] (0,0) -- (-0.5,-0.5);
            \draw[] (0,0) -- (0.6,0);
            \halfTenL{0,0}{ten1};
            \draw[loosely dotted, shorten <= 4pt, shorten >= 4pt] (-0.4,0.4) to[out=-110, in=110, looseness=1] (-0.4,-0.4);
        }{} 
        \ifthenelse{#1 = 6}{
            \draw[] (1.2,0) -- (1.7,0.5);
            \draw[] (1.2,0) -- (1.7,-0.5);
            \draw[] (0.6,0) -- (1.2,0);
            \ten{1.2,0}{ten1b};
            \halfTenR{0.7,0}{ten3};
            \draw[loosely dotted, shorten <= 4pt, shorten >= 4pt] (1.6,0.4) to[out=-70, in=70, looseness=1] (1.6,-0.4);
        }{}
        \ifthenelse{#1 = 7}{
            \draw[] (1.2,0) -- (1.7,0.5);
            \draw[] (1.2,0) -- (1.7,-0.5);
            \draw[] (0.6,0) -- (1.2,0);
            \halfTenR{1.2,0}{ten1};
            \draw[loosely dotted, shorten <= 4pt, shorten >= 4pt] (1.6,0.4) to[out=-70, in=70, looseness=1] (1.6,-0.4);
        }{}
        \ifthenelse{#1 = 8}{
            \draw[] (0,0) -- (-0.5,0.5);
            \draw[] (0,0) -- (-0.5,-0.5);
            \draw[] (0.6,0) -- (1.1,0.5);
            \draw[] (0.6,0) -- (1.1,-0.5);
            \draw[] (0,0) -- (0.6,0);
            \halfTenL{0,0}{ten1};
            \halfTenR{0.6,0}{ten1};
            \draw[loosely dotted, shorten <= 4pt, shorten >= 4pt] (-0.4,0.4) to[out=-110, in=110, looseness=1] (-0.4,-0.4);
            \draw[loosely dotted, shorten <= 4pt, shorten >= 4pt] (1,0.4) to[out=-70, in=70, looseness=1] (1,-0.4);
        }{}
	\end{tikzpicture}
}

\newcommand{\partitionFunction}[1]{
	\begin{tikzpicture}[scale=0.85,baseline={([yshift=-.5ex]current bounding box.center)}]
	    \coordinate (1) at (-0.5,0);
	    \coordinate (2) at (1.5,0);
	    \coordinate (3) at (-0.5,1);
	    \coordinate (4) at (1.5,1);
	    \coordinate (5) at (0,-0.5);
	    \coordinate (6) at (0,1.5);
	    \coordinate (7) at (1,-0.5);
	    \coordinate (8) at (1,1.5);
	    \coordinate (11) at (-0.5,0.3);
	    \coordinate (12) at (1.5,0.3);
	    \coordinate (13) at (-0.5,1.3);
	    \coordinate (14) at (1.5,1.3); 
	    \coordinate (15) at (0.3,-0.5);
	    \coordinate (16) at (0.3,1.5);
	    \coordinate (17) at (1.3,-0.5);
	    \coordinate (18) at (1.3,1.5);
	    \coordinate (20) at (0.3,0.5);
	    \coordinate (21) at (1.3,0.5);
	    
        \draw[] (1) -- (2);
        \draw[] (3) -- (4);
        \draw[] (5) -- (6);
        \draw[] (7) -- (8);
        \draw[] (1) to[out=180, in =180, looseness=2] (11);
        \draw[] (2) to[out=0, in =0, looseness=2] (12);
        \interLL{11}{12}{5}{6}{7}{8}{}
        \draw[] (3) to[out=180, in =180, looseness=2] (13);
        \draw[] (4) to[out=0, in =0, looseness=2] (14);
        \interLL{13}{14}{5}{6}{7}{8}{}
        \draw[] (5) to[out=-90, in =-90, looseness=2] (15);
        \draw[] (6) to[out=90, in=90, looseness=2] (16);
        \interLL{15}{20}{1}{2}{11}{12}{}
        \interLL{20}{16}{3}{4}{13}{14}{}
        \draw[] (7) to[out=-90, in =-90, looseness=2] (17);
        \draw[] (8) to[out=90, in=90, looseness=2] (18);
        \interLL{17}{21}{1}{2}{11}{12}{}
        \interLL{21}{18}{3}{4}{13}{14}{}
        \ifthenelse{#1 = 1}{
             \network{0}{0}{1}{1}{ten0}
        }{}
        \ifthenelse{#1 = 2}{
            \deltaTen{0,0};
            \deltaTen{0,1};
            \deltaTen{1,0};
            \deltaTen{1,1};
            \tenSmall{0}{0.5}{ten0c}{0};
            \tenSmall{1}{0.5}{ten0c}{0};
            \tenSmall{0.5}{0}{ten0c}{0};
            \tenSmall{0.5}{1}{ten0c}{0};
            \tenSmall{0.3}{-0.35}{ten0c}{0};
            \tenSmall{-0.35}{1.3}{ten0c}{0};
            \tenSmall{-0.35}{0.3}{ten0c}{0};
            \tenSmall{1.3}{-0.35}{ten0c}{0};
        }{}
	\end{tikzpicture}
}

\newcommand{\dummyContraction}[1]{
	\begin{tikzpicture}[scale=0.85,baseline={([yshift=-.5ex]current bounding box.center)}]
	\ifthenelse{#1 = 0}{
        \draw[] (-0.5,0) -- (0.5,0);
        \draw[] (0,0.5) -- (0,-0.5);
        \ten{0,0}{ten0};
        \node[above] at (-0.5,0) {${\sss \chi}$};
        \node[below] at (0.5,0) {${\sss \chi}$};
        \node[left] at (0,-0.5) {${\sss \chi}$};
        \node[right] at (0,0.5) {${\sss \chi}$};
	}{}
	\ifthenelse{#1 = 1}{
    	\begin{scope}
        	\clip (-0.5,-0.5) rectangle (1.5,1.5);
            \lattice{-1}{-1}{2}{2}	    
            \network{0}{0}{1}{1}{ten0}
    	\end{scope}
	}{}
	\ifthenelse{#1 = 2}{
        \draw[] (-1,0.03) -- (1,0.03);
        \draw[] (-1,-0.03) -- (1,-0.03);
        \draw[] (-0.03,1) -- (-0.03,-1);
        \draw[] (0.03,1) -- (0.03,-1);
        \tenBig{0,0}{ten0};
        \node[above] at (-1,0) {${\sss \chi^2}$};
        \node[below] at (1,0) {${\sss \chi^2}$};
        \node[left] at (0,-1) {${\sss \chi^2}$};
        \node[right] at (0,1) {${\sss \chi^2}$};
	}{}
	\end{tikzpicture}
}

\newcommand{\TRG}[1]{
	\begin{tikzpicture}[scale=0.85,baseline={([yshift=-.5ex]current bounding box.center)}]
	\ifthenelse{#1 = 0}{
    	\begin{scope}
        	\clip (-0.5,-0.5) rectangle (3.5,3.5);
            \lattice{-1}{-1}{4}{4}	    
            \network{0}{0}{3}{3}{ten0}
    	\end{scope}
	}{}
	\ifthenelse{#1 = 1}{
    	\begin{scope}
        	\clip (-0.5,-0.5) rectangle (3.5,3.5);
            \lattice{-1}{-1}{4}{4}	    
            \network{0}{0}{3}{3}{ten0}
    	\end{scope}
	}{}
	\ifthenelse{#1 = 2}{
    	\begin{scope}
        	\clip (-0.5,-0.5) rectangle (3.5,3.5);
        	\foreach \x in {0.5,2.5}
        	    \foreach \y in {-0.5,1.5,3.5}
                    \TRGUnitCell{\x}{\y};
            \foreach \x in {-0.5,1.5,3.5}
        	    \foreach \y in {0.5,2.5}
                    \TRGUnitCell{\x}{\y};          
    	\end{scope}
	}{}
	\ifthenelse{#1 = 3}{
    	\begin{scope}
        	\clip (-0.5,-0.5) rectangle (3.5,3.5);
            \draw[] (0.5,-0.5) -- (-0.5,0.5) -- (2.5,3.5) -- (3.5,2.5) -- (0.5,-0.5);
            \draw[] (-0.5,2.5) -- (0.5,3.5) -- (3.5,0.5) -- (2.5,-0.5) -- (-0.5,2.5);
            \foreach \x in {0.5,2.5}
                \foreach \y in {-0.5,1.5,3.5}
                    \tenBig{\x,\y}{ten0};
            \foreach \x in {-0.5,1.5,3.5}
                \foreach \y in {0.5,2.5}
                    \tenBig{\x,\y}{ten0};  
    	\end{scope}
	}{}
	\ifthenelse{#1 = 4}{
    	\begin{scope}
        	\clip (-0.5,-0.5) rectangle (3.5,3.5);
        	\foreach \x in {0.5,2.5}
        	    \foreach \y in {-0.5,1.5,3.5}
                    \TRGUnitCell{\x}{\y};
            \foreach \x in {-0.5,1.5,3.5}
        	    \foreach \y in {0.5,2.5}
                    \TRGUnitCell{\x}{\y};
            \foreach \x in {0,2,4}
                \foreach \y in {0,2,4} 
                    \CDL{\x-0.5,\y-0.5}{0.58}; 
            \foreach \x in {1,3}
                \foreach \y in {1,3} 
                    \CDL{\x-0.5,\y-0.5}{0.58}; 
            \foreach \x in {1,3}
                \foreach \y in {0,2,4} 
                    \CDL{\x-0.5,\y-0.5}{0.28}; 
            \foreach \x in {0,2,4}
                \foreach \y in {1,3} 
                    \CDL{\x-0.5,\y-0.5}{0.28};           
    	\end{scope}
	}{}
	\ifthenelse{#1 = 5}{
    	\begin{scope}
        	\clip (-0.5,-0.5) rectangle (3.5,3.5);
            \draw[] (0.5,-0.5) -- (-0.5,0.5) -- (2.5,3.5) -- (3.5,2.5) -- (0.5,-0.5);
            \draw[] (-0.5,2.5) -- (0.5,3.5) -- (3.5,0.5) -- (2.5,-0.5) -- (-0.5,2.5);
            \foreach \x in {0.5,2.5}
                \foreach \y in {-0.5,1.5,3.5}
                    \tenBig{\x,\y}{ten0};
            \foreach \x in {-0.5,1.5,3.5}
                \foreach \y in {0.5,2.5}
                    \tenBig{\x,\y}{ten0};  
            \foreach \x in {0,2,4}
                \foreach \y in {0,2,4} 
                    \CDL{\x-0.5,\y-0.5}{0.63}; 
            \foreach \x in {1,3}
                \foreach \y in {1,3} 
                    \CDL{\x-0.5,\y-0.5}{0.63}; 
    	\end{scope}
	}{}
	\end{tikzpicture}
}

\newcommand{\GiltOpt}[1]{
	\begin{tikzpicture}[scale=0.85,baseline=-0.3em]
        \draw[] (-1,0.5) -- (1,0.5);
        \draw[] (-1,-0.5) -- (1,-0.5);
        \draw[] (-0.5,1) -- (-0.5,-1);
        \draw[] (0.5,1) -- (0.5,-1);
        \ten{-0.5,-0.5}{ten0};
        \ten{0.5,-0.5}{ten0};
        
        \ifthenelse{#1 = 1}{
            \ten{-0.5,0.5}{ten0};
            \ten{0.5,0.5}{ten0};
            \node[above] at (0,0.5) {${\sss \chi}$};
        }{}
        \ifthenelse{#1 = 101}{
            \ten{-0.5,0.5}{ten0};
            \ten{0.5,0.5}{ten0};
        }{}
        \ifthenelse{#1 = 2}{
            \ten{-0.5,0.5}{ten0};
            \ten{0.5,0.5}{ten0};
            \tenSmall{0}{0.5}{ten4}{2};
            \node[above] at (-0.25,0.5) {${\sss \chi}$};
            \node[above] at (0.25,0.5) {${\sss \chi}$};
        }{}
        \ifthenelse{#1 = 3}{
            \ten{-0.5,0.5}{ten0};
            \ten{0.5,0.5}{ten0};
            \halfTenSmallL{-0.08}{0.5}{ten4}{2};
            \halfTenSmallR{0.08}{0.5}{ten4}{2};
            \node[above, yshift=2pt] at (0,0.5) {${\sss \;\, \chi'}$};
        }{}
        \ifthenelse{#1 = 4}{
            \centerarc[tenBG](-0.5,0.5)(0:360:0.16)
            \centerarc[fill = BlueViolet, fill opacity=0.4](-0.5,0.5)(0:360:0.16)
            \centerarc[tenBG](0.5,0.5)(0:360:0.16)
            \centerarc[fill = BlueViolet, fill opacity=0.4](0.5,0.5)(0:360:0.16)
            \begin{scope}
                \clip (-0.5,{0.5-0.16}) arc (-90:90:0.16);
                \fill[white] ({-0.5+0.16},{0.5+0.16}) -- (-0.5,0.5) -- ({-0.5+0.16},{0.5-0.16}) -- ({-0.5+0.16},{0.5+0.16});
                \draw[draw=Thistle, fill=Thistle, opacity=0.5] ({-0.5+0.16},{0.5+0.16}) -- (-0.5,0.5) -- ({-0.5+0.16},{0.5-0.16}) -- ({-0.5+0.16},{0.5+0.16});
            \end{scope}
            \begin{scope}
                \clip (0.5,{0.5+0.16}) arc (90:270:0.16);
                \fill[white] ({0.5-0.16},{0.5+0.16}) -- (0.5,0.5) -- ({0.5-0.16},{0.5-0.16}) -- ({0.5-0.16},{0.5+0.16});
                \draw[draw=NavyBlue, fill=NavyBlue, opacity=0.35] ({0.5-0.16},{0.5+0.16}) -- (0.5,0.5) -- ({0.5-0.16},{0.5-0.16}) -- ({0.5-0.16},{0.5+0.16});
            \end{scope}
            \centerarc[draw=black](-0.5,0.5)(0:360:0.16)
            \centerarc[draw=black](0.5,0.5)(0:360:0.16)
            \node[above] at (0,0.5) {${\sss \;\, \chi'}$};
        }{}
        \ifthenelse{#1 = 5}{
            \ten{-0.5,0.5}{ten0};
            \ten{0.5,0.5}{ten0};
            \draw[] (0,{0.5+0.18}) -- (0,{1.2-0.18});
            \centerarc[ten1](0,0.5)(0:180:0.18)
            \centerarc[ten1](0,1.2)(0:-180:0.18)
            \draw[] (-0.18,1.2) -- (0.18,1.2);
            \draw[] (-0.18,1.2) to[out=90, in=90, looseness=2] (0.18,1.2);
            \node[right, xshift=2pt] at (0,1.2) {$U$};
            \node[below] at (0,0.5) {$U^\dagger$};
        }{}
        \ifthenelse{#1 = 6}{
            \ten{-0.5,0.5}{ten0};
            \ten{0.5,0.5}{ten0};
            \draw[] (0,{0.5+0.18}) -- (0,{1.2-0.12});
            \centerarc[ten1](0,0.5)(0:180:0.18)
            \centerarc[ten1](0,1.2)(0:360:0.12)
            \node[right] at (0,1.2) {$t$};
        }{}
	\end{tikzpicture}
}

\newcommand{\GiltOptSVD}[1]{
	\begin{tikzpicture}[scale=0.85,baseline=-0.3em]
        \ifthenelse{#1 = 1}{
            \draw[] (-1,0.5) -- (-0.2,0.5);
            \draw[] (0.2,0.5) -- (1,0.5);
            \draw[] (-1,-0.5) -- (1,-0.5);
            \draw[] (-0.5,1) -- (-0.5,-1);
            \draw[] (0.5,1) -- (0.5,-1);
            \ten{-0.5,-0.5}{ten0};
            \ten{0.5,-0.5}{ten0};
            \ten{-0.5,0.5}{ten0};
            \ten{0.5,0.5}{ten0};
        }{}
        \ifthenelse{#1 = 2}{
            \draw[] (-1,0.5) -- (-0.5,0.5);
            \draw[] (-0.5,0.5) to[out=0, in =-90, looseness=1.5] (-0.18,1);
            \draw[] (0.5,0.5) -- (1,0.5);
            \draw[] (0.5,0.5) to[out=180, in =-90, looseness=1.5] (0.18,1);
            \draw[] (-1,-0.5) -- (1,-0.5);
            \draw[] (-0.5,1) -- (-0.5,-1);
            \draw[] (0.5,1) -- (0.5,-1);
            \ten{-0.5,-0.5}{ten0};
            \ten{0.5,-0.5}{ten0};
            \ten{-0.5,0.5}{ten0};
            \ten{0.5,0.5}{ten0};
            \draw[dotted, draw=black, fill = BlueViolet, fill opacity=0.1] (-0.75,0.5) to[out=90, in=180, looseness=1] (-0.5,0.75) to[out=0, in=90, looseness=1] (-0.25,0.5) -- (-0.25,0) to[out=-90, in =180, looseness=1] (0,-0.25) to[out=0, in=-90, looseness=1] (0.25,0) -- (0.25,0.5) to[out=90, in=180, looseness=1] (0.5,0.75) to[out=0, in=90, looseness=1] (0.75,0.5) -- (0.75,-0.5) to[out=-90, in=0, looseness=1] (0.5,-0.75) -- (-0.5,-0.75) to[out=180,in=-90, looseness=1] (-0.75,-0.5) -- (-0.75,0.5);
        }{}
        \ifthenelse{#1 = 3}{
            \draw[] (-1,0.5) -- (-0.5,0.5);
            \draw[] (0.5,0.5) -- (1,0.5);
            \draw[] (-1,-0.5) -- (1,-0.5);
            \draw[] (-0.5,1) -- (-0.5,-1);
            \draw[] (0.5,1) -- (0.5,-1);
            \fill[white] (-0.75,0.5) to[out=90, in=180, looseness=1] (-0.5,0.75) to[out=0, in=90, looseness=1] (-0.25,0.5) -- (-0.25,0) to[out=-90, in =180, looseness=1] (0,-0.25) to[out=0, in=-90, looseness=1] (0.25,0) -- (0.25,0.5) to[out=90, in=180, looseness=1] (0.5,0.75) to[out=0, in=90, looseness=1] (0.75,0.5) -- (0.75,-0.5) to[out=-90, in=0, looseness=1] (0.5,-0.75) -- (-0.5,-0.75) to[out=180,in=-90, looseness=1] (-0.75,-0.5) -- (-0.75,0.5);
            \draw[ten1] (-0.75,0.5) to[out=90, in=180, looseness=1] (-0.5,0.75) to[out=0, in=90, looseness=1] (-0.25,0.5) -- (-0.25,0) to[out=-90, in =180, looseness=1] (0,-0.25) to[out=0, in=-90, looseness=1] (0.25,0) -- (0.25,0.5) to[out=90, in=180, looseness=1] (0.5,0.75) to[out=0, in=90, looseness=1] (0.75,0.5) -- (0.75,-0.5) to[out=-90, in=0, looseness=1] (0.5,-0.75) -- (-0.5,-0.75) to[out=180,in=-90, looseness=1] (-0.75,-0.5) -- (-0.75,0.5);
            \draw[] (-0.18,1.7) -- (-0.18,1.5);
            \draw[] (0.18,1.7) -- (0.18,1.5);
            \draw[] (0,{1.5-0.18}) -- (0,-0.25);
            \centerarc[ten1](0,1.5)(0:-180:0.18)
            \draw[] (-0.18,1.5) -- (0.18,1.5);
            \ten{0,0.5}{ten3};
            \node[right, xshift=2pt] at (0,1.5) {$U$};
            \node[right] at (0.75,0) {$V^\dagger$};
            \node[left, xshift=2.5pt] at (0,0.9) {$S$};
        }{}
	\end{tikzpicture}
}

\newcommand{\GiltMatrix}[1]{
	\begin{tikzpicture}[scale=0.85,baseline=-0.3em]
        \draw[] (-0.5,0) -- (0.5,0);
        \ifthenelse{#1 = 1}{
            \tenSmall{0}{0}{ten4}{2};
        }{}
        \ifthenelse{#1 = 2}{
            \draw[] (0,{0.18}) -- (0,{0.7-0.12});
            \centerarc[ten1](0,0)(0:180:0.18)
            \centerarc[ten4](0,0.7)(0:360:0.12)
            \node[right] at (0,0.7) {$t'$};
        }{}
	\end{tikzpicture}
}

\newcommand{\TNR}[1]{
	\begin{tikzpicture}[scale=0.85,baseline={([yshift=-.5ex]current bounding box.center)}]
	    \begin{scope}
            \clip (-0.5,-0.5) rectangle (3.5,3.5);
            
            \ifthenelse{#1 = 1}{
                \lattice{-1}{-1}{4}{4}	    
                \network{0}{0}{3}{3}{ten0}
                \foreach \x in {0,...,4}
                    \foreach \y in {0,...,4} 
                        \CDL{\x-0.5,\y-0.5}{0.43};
                \foreach \x in {0.5,2.5}
                    \foreach \y in {0.5,2.5}
                        \TNRUnitCell{\x}{\y}{1};
                \foreach \x in {-0.5,1.5,3.5}
                    \foreach \y in {-0.5,1.5,3.5}
                        \TNRUnitCell{\x}{\y}{1};  
            }{}
            
            \ifthenelse{#1 = 2}{
                \lattice{-1}{-1}{4}{4}	    
                \network{0}{0}{3}{3}{ten0}
                \foreach \x in {-0.5,1.5,3.5}
                    \foreach \y in {-0.5,1.5,3.5}
                        \brokenCDL{\x,\y}{0.43};
                \foreach \x in {0.5,2.5}
                    \foreach \y in {0.5,2.5}
                        \brokenCDL{\x,\y}{0.43};
                \foreach \x in {0.5,2.5}
                    \foreach \y in {-0.5,1.5,3.5}
                        \CDL{\x,\y}{0.43};
                \foreach \x in {-0.5,1.5,3.5}
                    \foreach \y in {0.5,2.5}
                        \CDL{\x,\y}{0.43};
                \foreach \x in {0.5,2.5}
                    \foreach \y in {0.5,2.5}
                        \TNRUnitCell{\x}{\y}{2}; 
                \foreach \x in {-0.5,1.5,3.5}
                    \foreach \y in {-0.5,1.5,3.5}
                        \TNRUnitCell{\x}{\y}{2}; 
            }{}
            
            \ifthenelse{#1 = 3}{
                \lattice{-1}{-1}{4}{4}	    
                \networkBi{ten2}{ten5}
                \foreach \x in {0.5,2.5}
                    \foreach \y in {-0.5,1.5,3.5}
                        \CDL{\x,\y}{0.43};
                \foreach \x in {-0.5,1.5,3.5}
                    \foreach \y in {0.5,2.5}
                        \CDL{\x,\y}{0.43};
            }{}
            
            \ifthenelse{#1 = 4}{
                \foreach \x in {0.5,2.5}
            	    \foreach \y in {-0.5,1.5,3.5}
                        \TRGUnitCellBi{\x}{\y}{ten2}{ten5};
                \foreach \x in {-0.5,1.5,3.5}
            	    \foreach \y in {0.5,2.5}
                        \TRGUnitCellBi{\x}{\y}{ten2}{ten5}; 
                \foreach \x in {1,3}
                    \foreach \y in {0,2,4} 
                        \CDL{\x-0.5,\y-0.5}{0.28}; 
                \foreach \x in {0,2,4}
                    \foreach \y in {1,3} 
                        \CDL{\x-0.5,\y-0.5}{0.28};         
            }{}
    	\end{scope}
	\end{tikzpicture}
}

\newcommand{\miniTNR}[1]{
	\begin{tikzpicture}[scale=0.85,baseline={([yshift=-.5ex]current bounding box.center)}]
	    \begin{scope}
            \clip (-0.5,-0.5) rectangle (1.5,1.5);
            \ifthenelse{#1 = 0}{
                \lattice{-1}{-1}{4}{4}	    
                \network{0}{0}{3}{3}{ten0}
                \foreach \x in {-0.5,0.5,1.5}
                    \foreach \y in {-0.5,0.5,1.5}
                        \CDL{\x,\y}{0.43}; 
            }{}
            \ifthenelse{#1 = 1}{
                \lattice{-1}{-1}{4}{4}	    
                \network{0}{0}{3}{3}{ten0}
                \foreach \x in {-0.5,0.5,1.5}
                    \foreach \y in {-0.5,0.5,1.5}
                        \CDL{\x,\y}{0.43};  
                \foreach \x in {0.5,2.5}
                    \foreach \y in {0.5,2.5}
                        \TNRUnitCell{\x}{\y}{1};
                \foreach \x in {-0.5,1.5,3.5}
                    \foreach \y in {-0.5,1.5,3.5}
                        \TNRUnitCell{\x}{\y}{1};          
            }{}
            \ifthenelse{#1 = 2}{
                \lattice{-1}{-1}{4}{4}	    
                \network{0}{0}{3}{3}{ten0}
                \brokenCDL{-0.5,-0.5}{0.43}; 
                \brokenCDL{1.5,-0.5}{0.43}; 
                \brokenCDL{0.5,0.5}{0.43}; 
                \brokenCDL{-0.5,1.5}{0.43}; 
                \brokenCDL{1.5,1.5}{0.43}; 
                \CDL{0.5,-0.5}{0.43};
                \CDL{-0.5,0.5}{0.43};
                \CDL{1.5,0.5}{0.43};
                \CDL{0.5,1.5}{0.43};
                \foreach \x in {0.5,2.5}
                    \foreach \y in {0.5,2.5}
                        \TNRUnitCell{\x}{\y}{2}; 
                \foreach \x in {-0.5,1.5,3.5}
                    \foreach \y in {-0.5,1.5,3.5}
                        \TNRUnitCell{\x}{\y}{2}; 
            }{}
            \ifthenelse{#1 = 3}{
                \lattice{-1}{-1}{4}{4}	    
                \networkBi{ten2}{ten5}
                \CDL{0.5,-0.5}{0.43};
                \CDL{-0.5,0.5}{0.43};
                \CDL{1.5,0.5}{0.43};
                \CDL{0.5,1.5}{0.43};
            }{}
    	\end{scope}
	\end{tikzpicture}
}

\newcommand{\GiltTensor}[1]{
	\begin{tikzpicture}[scale=0.85,baseline={([yshift=-.5ex]current bounding box.center)}]
        \draw[] (-0.5,0) -- (0.5,0);
        \draw[] (0,-0.5) -- (0,0.5);
        \ifthenelse{#1 = 1}{
            \ten{0,0}{ten5};
        }{}
    	\ifthenelse{#1 = 2}{
            \ten{0,0}{ten0};
            \halfTenSmallB{0}{0.5-0.08}{ten4}{1};
            \halfTenSmallT{0}{-0.5+0.08}{ten4}{4};
            \halfTenSmallL{0.5-0.08}{0}{ten4}{3};
            \halfTenSmallR{-0.5+0.08}{0}{ten4}{2};
    	}{}
    	\ifthenelse{#1 = 3}{
            \ten{0,0}{ten2};
        }{}
    	\ifthenelse{#1 = 4}{
            \draw[] (-0.5,0) -- (0.5,0);
            \draw[] (0,-0.5) -- (0,0.5);
            \ten{0,0}{ten0};
            \halfTenSmallB{0}{0.5-0.08}{ten4}{4};
            \halfTenSmallT{0}{-0.5+0.08}{ten4}{1};
            \halfTenSmallL{0.5-0.08}{0}{ten4}{2};
            \halfTenSmallR{-0.5+0.08}{0}{ten4}{3};
    	}{}
	\end{tikzpicture}
}

\newcommand{\spectrum}[1]{
	\begin{tikzpicture}[scale=0.85,baseline={([yshift=-.5ex]current bounding box.center)}]
	\ifthenelse{#1 = 1}{
        \draw[] (-0.5,0) -- (0.5,0);
        \draw[] (0,0.5) -- (0,-0.5);
        \ten{0,0}{ten0};
	}{}
	\ifthenelse{#1 = 2}{
	    \draw[] (-0.4,0.9) -- (-0.4,0.4) -- (-0.9,0.4);
	    \draw[] (0.4,-0.9) -- (0.4,-0.4) -- (0.9,-0.4);
	    \draw[] (-0.4,0.4) -- (0.4,-0.4);
	    \ten{-0.4,0.4}{ten0d};
        \ten{0,0}{ten3};
        \ten{0.4,-0.4}{ten0d};
        \node[right, xshift=0pt, yshift=6pt] at (0,0) {$S$};
	}{}
	\end{tikzpicture}
}

\newcommand{\transfer}[0]{
	\begin{tikzpicture}[scale=0.85,baseline={([yshift=-.5ex]current bounding box.center)}]
	    \coordinate (2) at (-0.5,0);
	    \coordinate (3) at (1.5,0);
	    \coordinate (4) at (-0.5,0.3);
	    \coordinate (5) at (1.5,0.3);
	    \coordinate (6) at (0,-0.5);
	    \coordinate (7) at (0,0.5);
	    \coordinate (8) at (1,-0.5);
	    \coordinate (9) at (1,0.5);
        \draw[] (2) -- (3);
        \draw[] (2) to[out=180, in=180, looseness=2] (4);
        \draw[] (3) to[out=0, in=0, looseness=2] (5);
        \draw[] (6) -- (7);
        \draw[] (8) -- (9);
        \interLL{4}{5}{6}{7}{8}{9}{};
        \ten{0,0}{ten0};
        \ten{1,0}{ten0};
	\end{tikzpicture}
}

\newcommand{\specDummyLines}[1]{
    \path[name path=P1] (1.7,0.5) .. controls ++(-0.1,2) .. (sp2);
    \path[name path=P2]  (1.9,0.5) .. controls ++(0.1,1.8) .. (sp2);
    \path[name path=P3]  (2.1,0.5) .. controls ++(0.3,1.6) .. (sp2);
    \path[name path=C] (1,{#1+0.5}) -- (6,{#1+0.8});
    \path [name intersections={of=P1 and C,by={a}}];
    \path [name intersections={of=P2 and C,by={b}}];
    \path [name intersections={of=P3 and C,by={c}}];
    \node[specCirc] at (a) {};
    \node[specCirc] at (b) {};
    \node[specCirc] at (c) {};
}

\newcommand{\heuristicFlow}[0]{
	\begin{tikzpicture}[scale=0.83,baseline=-4.6em]
        \draw[line width=0.8pt] (0.2,-0.5) rectangle (5.8,4.5);
        \draw[->1-=0.25] (1.5,0.5) to[out=90, in=-45, looseness=1, edge node={coordinate[pos=0.4] (sp1)}] (0.5,4);
        \draw[->1-=0.25] (0.5,4) to[in=90, out=-45, looseness=1] (1.5,0.5);
        \path (5.5,1) to[out=90, in=-45, looseness=1, edge node={coordinate[pos=0.6] (sp2)}] (4.5,4.5);
        \node[circle, fill=black, inner sep=1.5pt] at (sp1) {};
        \node[circle, fill=black, inner sep=1.5pt] at (sp2) {};
        \draw[->1-=0.55] (sp1) to[out=40, in=185, looseness=1] (sp2);

        \draw[dotted, color=BlueViolet, line width=0.6pt] (1.7,0.5) .. controls ++(-0.1,2) .. (sp2) node[pos=0.5] (sp4) {};
        \draw[dotted, color=BlueViolet, line width=0.6pt]  (1.9,0.5) .. controls ++(0.1,1.8) .. (sp2) node[pos=0.6] (sp5) {};
        \draw[dotted, color=BlueViolet, line width=0.6pt]  (2.1,0.5) .. controls ++(0.3,1.6) .. (sp2) node[pos=0.73] (sp6) {};
        
        \foreach \x in {0,0.6,1.1,1.5,1.8,2,2.15,2.25,2.32,2.37,2.4,2.42}
            \specDummyLines{\x};
        \node[left, xshift=2pt, yshift=-6pt] at (sp1)  {$\text{\parbox[t]{1cm}{\small QFT UV}}$}; 
        \node[above, yshift=3pt, xshift=-1pt] at (sp2) {$\text{\parbox[t]{1cm}{\small QFT IR}}$}; 
        \node[below] at (1.8,0.5) {$\text{\parbox[t]{1.5cm}{\small Lattice UV}}$}; 
        \node[] at (4,1) {$\text{\small $\lambda < \lambda ' < \lambda''$}$};
        \draw[line width=0.3pt, ->] (3.1,1.2) to[out=90, in=-10, looseness=1] (sp4);
        \draw[line width=0.3pt, ->] (3.9,1.2) to[out=90, in=-10, looseness=1] (sp5);
        \draw[line width=0.3pt, ->] (4.7,1.2) to[out=90, in=-10, looseness=1] (sp6);
	\end{tikzpicture}
}

\hypersetup{%
	bookmarksopen=true,
	unicode=true,           
	pdftoolbar=true,        
	pdfmenubar=true,        
	pdffitwindow=false,     
	pdfstartview={FitH},    
	pdftitle={Computing the renormalization group flow of two-dimensional \texorpdfstring{$\bm{\phi^4}$}{phi\^ 4} theory with tensor networks},
	pdfauthor={},
	pdfsubject={},          
	pdfcreator={},          
	pdfproducer={pdfLaTeX}, 
	pdfkeywords={Gilt} {quantum field theory} {tensor networks} {renormalization},
	pdfnewwindow=true,      
	colorlinks,
	citecolor=BlueViolet,
	filecolor=BlueViolet,
	linkcolor=BlueViolet,
	urlcolor=BlueViolet,
	linktocpage=true
}

\definecolor{cbl}{rgb}{0,0,1}
 
\definecolor{crd}{rgb}{1,0,0}
 
\newcommand{\upd}{\mathrm{d}}

\newcommand{\ie}[0]{\textit{i.e.} }
\newcommand{\eg}[0]{\textit{e.g.} }

\newcommand\e{\mathrm{e}}

\newcommand{\q}{\quad}
\newcommand{\sss}{\scriptstyle}
\newcommand{\nn}{\nonumber}
\newcommand{\fc}{11.0861(90)}

\begin{document}
\title{Computing the renormalization group flow of two-dimensional \texorpdfstring{$\bm{\phi^4}$}{phi\^ 4} theory \\ with tensor networks}
\author{\textsc{Clement Delcamp}}
\email{clement.delcamp@mpq.mpg.de}
\author{\textsc{Antoine Tilloy}}
\email{antoine.tilloy@mpq.mpg.de}
\affiliation{\vspace{1em}Max-Planck-Institut f\"ur Quantenoptik, Hans-Kopfermann-Stra{\ss}e 1, 85748 Garching, Germany}
\affiliation{Munich Center for Quantum Science and Technology (MCQST) Schellingstr. 4, D-80799 M{\"u}nchen, Germany}

\begin{abstract}
\noindent We study the renormalization group flow of $\phi^4$ theory in two dimensions. Regularizing space into a fine-grained lattice and discretizing the scalar field in a controlled way, we rewrite the partition function of the theory as a tensor network. Combining local truncations and a standard coarse-graining scheme, we obtain the renormalization group flow of the theory as a map in a space of tensors. Aside from qualitative insights, we verify the scaling dimensions at criticality and extrapolate the critical coupling constant $f_{\rm c} = \lambda  / \mu ^2$ to the continuum to find $f^{\rm cont.}_{\rm c} = \fc$, which favorably compares with alternative methods.
\end{abstract}

\maketitle

\section{Introduction}

\noindent
Solving interacting quantum field theories (QFTs) out of the perturbative regime is difficult, and is arguably one of the most pressing computational problems in theoretical physics nowadays. Among all non-trivial quantum field theories, the simplest one is perhaps the self-interacting scalar field in two spacetime dimensions, \emph{a.k.a.} $\phi^4_{2}$. This theory provides a good case study because it is reasonably simple to define (it is super-renormalizable), and was actually one of the first interacting QFTs to be rigorously constructed \cite{glimm1987,fernandez1992}. However, it does not bypass the real difficulties of QFT, namely it is neither free, nor integrable, nor supersymmetric. Furthermore, its physical interpretation is rather simple as it does not display multiple instantons or topological complications. Nevertheless, it does contain a second order phase transition, which is ubiquitous in physics. Finally, and perhaps most importantly, there are still some questions for which we do not possess exact answers.

Recently, there has been a renewal of interest for deterministic methods that aim at probing the deep non-perturbative regime of QFTs, and of $\phi^4_{\rm 2}$ in particular. One can mention Hamiltonian truncation techniques as well as their renormalized refinements  \cite{rychkov2015,eliasmiro2017-1,eliasmiro2017-2}, Borel resummation \cite{serone2018}, or tensor network based approaches \cite{milsted2013,vanhecke2019,kadoh2019}. Ultimately, one common objective of all these methods is to outperform lattice Monte-Carlo techniques, which are primarily used in non-perturbative high-energy computations, \eg in quantum chromodynamics \cite{kogut1983,muroya2003,fodor2012}. Although some of these new techniques temporarily outperformed the previous state of the art in the case of $\phi^4_{2}$, Monte Carlo methods \cite{schaich2009,bosetti2015} were quickly improved and now seem back on a par \cite{bronzin2019}. As part of this effort, we propose in this manuscript a computation of the renormalization group (RG) flow of $\phi^4_{2}$ using tensor network renormalization techniques. 

Tensor network methods, which have proven a very potent tool in the context of quantum information and condensed matter theory, can be  divided into two distinct families. In the first, one writes an ansatz for a low-energy state of a given Hamiltonian as a tensor network, and optimizes its parameters variationally so as to get closer to the true ground state \cite{PhysRevLett.69.2863, Michalakis_2006, Kl_mper_1993, Verstraete_2004, Daley_2004, Stoudenmire_2012, verstraete2004renormalization, Vidal_2008, Shi_2006}. In the second, one decomposes a Euclidean partition function directly as a tensor network, and the challenge then becomes to contract it approximately \cite{Levin:2006jai, Gu_2009, Xie_2009, Evenbly_2015, Yang_2017, Bal_2017, hauru2018}, with the aid of techniques akin to Kadanoff's blocking scheme \cite{PhysRevLett.34.1005}.  In both cases, and as the name suggests, tensor network methods have been primarily developed for lattice systems. To deal with QFTs, two options are available: Either one extends the tensor networks themselves to the continuum, as has been tried in recent years \cite{verstraete2010,haegeman2013_cMERA,haegeman2013_cMPS,tilloy2019}, or one discretizes the theory on a lattice, making the well-established tensor network techniques available. Although the first option is a promising initiative, we shall explore the second one in this manuscript. In short, we wish to study $\phi^4_{2}$ by discretizing its functional integral on a very fine-grained lattice, and solving it with tensor network renormalization techniques.

The best-known method for approximately contracting tensor networks that represent partition functions is the so-called Tensor Renormalization Group (\textsf{TRG}) algorithm. It was applied to $\phi^4_{2}$ in \cite{shimizu2012analysis}, and more recently in \cite{kadoh2019}, where the authors report an estimate of the critical coupling of the theory in the continuum limit. Although it is a simple and fairly efficient algorithm, it has been known for several years that, despite its name, \textsf{TRG} does not in fact yield a \emph{proper} renormalization group flow. Indeed, it was shown to violate a fundamental principle of RG, namely that all ultraviolet (UV) details relative to a scale below the one considered should be integrated away. Concretely, this means that the fixed points of the flow depend on non-universal features. This makes the identification of critical points more difficult, and the scale invariance taking place at conformal fixed points nearly impossible to witness. 

Over the years, several more advanced algorithms that address the shortcomings of \textsf{TRG}, have been proposed \cite{Evenbly_2015, Yang_2017, Bal_2017, hauru2018}.  In this paper, we use the algorithm that was last introduced, namely $\textsf{Gilt-TNR}$, which is based on the notion of \emph{graph independent local truncations}. In addition to being physically sounder than \textsf{TRG}, it yields notably more accurate observables. Another technical contribution of our study is the introduction of a more transparent and controlled field discretization.

Our main result is  the computation of the critical value of the coupling in the continuum limit. As a benchmark, we also compute the scaling dimensions of the theory, which are known analytically from conformal field theory. Apart from the arguable gain of precision brought by tensor network renormalization techniques, we shall emphasize how these methods provide some precious qualitative insight about the RG flow of the theory. Furthermore, this study permits us to highlight some interesting specificities of the tensor network approach. In particular, we will show that they are rather complementary to other approaches: They work well at strong coupling, with excellent behavior in the infrared (IR), but rather suffer in the UV if the fixed point is the massless free boson, which is the case for super-renormalizable and asymptotically free bosonic QFTs.

\bigskip \noindent
In sec.~\ref{sec:model} we recall the continuous and discrete formulations of the theory. The tensor network reformulation is explained in sec.~\ref{sec:TNFormulation}. In sec.~\ref{sec:Gilt} we present the \textsf{Gilt}-\textsf{TNR} algorithm as a tool to perform renormalization group transformation for tensor networks, and the method pursued to identify the critical coupling. The algorithm is applied to $\phi^4_{2}$ and the results are tabulated in sec.~\ref{sec:results}. Finally, in sec.~\ref{sec:discussion}, we discuss the possible source of errors, as well as generalizations to higher dimensions.

\section{The model\label{sec:model}}

\noindent
\emph{In this section, we define $\phi^4_{2}$ theory on the lattice and in the continuum, and recall some of its basic properties.}

\subsection{From the continuum to the lattice}

\noindent
At a rather heuristic level, one can see (Euclidean) $\phi^4_{2}$ theory as the theory of a random scalar field $\phi$ of probability measure
\begin{equation}
    P(\phi) =\frac{\exp(-S_E(\phi))}{\mathcal{Z}} \, ,
\end{equation}
with
\begin{equation}
    S_E(\phi) = \int_{\mathbb{R}^2} \frac{1}{2}(\nabla\phi)^2 + \frac{m_0^2}{2}\, \phi^2 + \frac{g}{4}\, \phi^4
\end{equation}
and $\mathcal{Z}=\int \mathcal{D}\phi \exp(-S_E(\phi))$.
Of course, taken at face value, this probability measure is ill-defined. One option forward is to regularize the theory by discretizing the continuum $\mathbb{R}^2$ into a lattice, localize the field on its vertices, and write
\begin{equation}
    S^\mathsf{a}_E(\phi)=\sum_{\langle i,j \rangle} \frac{(\phi_i-\phi_j)^2}{2} + \sum_{i} \frac{\mu_0^2}{2}\, \phi_i^2 + \frac{\lambda}{4}\,\phi_i^4 \, ,
\end{equation}
where $\mathsf{a}$ is the lattice unit length, $\langle i,j \rangle$ denotes a sum on nearest neighbors and
\begin{equation}\label{eq:naivescaling}
    \mu_0= \mathsf{a} m_0 ~~ \text{and} ~~ \lambda = \mathsf{a}^2 g \, .
\end{equation}
Naively, if there were no quartic term, taking the continuum limit would simply amount to sending $\mathsf a \to  0$, while keeping the dimensionful parameter $m_0$ fixed. However, it is well understood that taking the continuum limit in the presence of non-quadratic terms requires a more subtle scaling of the parameters of the lattice theory.

\subsection{Perturbative renormalization}

\noindent
To take the continuum limit, one has to take into account the fact that the leading effect of the quartic term is to shift the quadratic (mass) part by a term $\propto -\log(\mathsf a)$. Unless this effect is countered, the continuum limit is a trivial theory with an infinite mass.

One option is to add a mass counterterm and define
\begin{equation}
    \mu_0^2 :=\mu^2 - 3\lambda A(\mu^2) \, ,
\end{equation}
where $A(\mu^2)$ corresponds to the (lattice) tadpole diagram responsible for the logarithmic divergence:
\begin{align}
    A(\mu^2)&=    \tadpole{}
    \\ \nn
    &= \frac{1}{N^2}\! \sum_{k_1,k_2=0}^{N-1}\frac{1}{\mu^2 + 4 \left[\sin^2(\pi k_1/N) + \sin^2(\pi k_2/N)\right]} 
    \\ \nn
    &\;\; \underset{N\rightarrow + \infty}{\longrightarrow} \int_{[0,1]^2} \frac{\upd k_1 \upd k_2}{\mu^2 + 4 \left[\sin^2(\pi k_1)) + \sin^2(\pi k_2)\right]} \, . 
\end{align}
It can be shown that this prescription is equivalent to the normal ordering used in the operator representation~\cite{loinaz1998}. The continuum limit is then obtained by sending $\mathsf a$ (or equivalently $\lambda$) to $0$, while keeping $f=\lambda/\mu^2$ fixed. In  the case of $\phi^4_{2}$, this simple mass renormalization is the only requirement to make the continuum limit well defined.

\subsection{Universal and non-universal properties}

\noindent
Let us now summarize what is known about the behavior of $\phi^4_{2}$, and distinguish its universal and non-universal properties.

 First, the lattice theory has two phases separated by a second order phase transition along a line in $(\lambda,\mu)$ space. For a fixed $\lambda$, this phase transition is reached for a value $\mu_{\rm c}(\lambda) \neq 0$. Whereas $\mu$ is a renormalized mass, since it contains a counter-term necessary to make the theory well defined in the continuum limit, it is not the physical mass associated with the spectral gap, which is a fully non-perturbative quantity. The continuum field theory also has a phase transition for  $f^\text{cont.}_{\rm c}:=\lim_{\lambda\rightarrow 0}\lambda/\mu^2_{\rm c}(\lambda) \equiv g/m_{\rm c}^2$. Since $\phi^4_{2}$ is a super-renormalizable QFT, defining its continuum limit is straightforward. It is a UV problem that is weakly coupled, and thus perturbatively solvable -- only one loop is required.
 However, finding the position of the critical point, in the discrete setting as well as in the continuum theory, is a difficult problem. Indeed, it is an IR problem, and thus strongly coupled and non-perturbative \cite{fernandez1992}. The most recent Monte-Carlo study \cite{bronzin2019} provides the estimate $f^\text{cont.}_{\rm c}\simeq 11.055(20)$, which is more accurate than Borel-resumming the perturbation expansion to $8^{\rm th}$ order in $\lambda$ \cite{serone2018}.

That being said, the IR properties of $\phi^4_{2}$ exactly at criticality are known analytically from conformal field theory (CFT). The critical point is in the Ising universality class, and is the same on the whole $(\lambda,\mu_{\rm c}(\lambda))$-line. Computing precisely the scaling dimensions on the critical line thus yields no new knowledge, but provides a good sanity check for any method aiming to solve $\phi^4_{2}$.

\section{Tensor network reformulation\label{sec:TNFormulation}}

\noindent
\emph{In this section, we explain how the partition function of the discretized theory can be computed as the contraction of a discrete tensor network.}

\subsection{Tensor network with field indices}

\noindent
Given the discretized action $S^\mathsf{a}_E$, we consider the lattice partition function
\begin{equation}
    \mathcal{Z}^{\mathsf a} = \int \prod_{i}\upd \phi_i \, \exp\left[-S^\mathsf{a}_E(\phi)\right] \, .
\end{equation}
Assuming that we accept continuous indices, this partition function can be rewritten as a tensor network \cite{kadoh2019}, which in the case of a $2 \times 2$ lattice reads
\begin{equation}
    \mathcal{Z}^{\mathsf a}_{2 \times 2} = \partitionFunction{2} \, ,
\end{equation}
where we introduced the tensors
\begin{align}
    &F(\phi_\ell,\phi_r) \equiv \matrixInit{1} \\
    \nn  & = \exp \Big[- \frac{1}{2}(\phi_\ell-\phi_r)^2-\frac{\mu^2_0}{8}(\phi_\ell^2 + \phi_r^2) - \frac{\lambda}{16}(\phi_\ell^4 + \phi_r^4)\Big] 
\end{align}
\begin{equation*}
    {\rm and }\;\;\; \deltaInit  \equiv \delta(\phi_\ell-\phi_u)\delta(\phi_u-\phi_r)\delta(\phi_r-\phi_d) \,.
\end{equation*}
In these diagrams, a connected link means a ``continuous contraction'', \ie an integral $\int_\mathbb{R}\upd \phi$ over real numbers. 
Equivalently, one could eliminate the Dirac deltas and define that each vertex corresponds to a field integral.

\subsection{Field discretization}

\noindent
To make the previous tensor network with continuous indices amenable to standard techniques, it is necessary to discretize the field variable. One could simply digitize the field $\phi$ \cite{shimizu2012analysis}, or use Gauss-Hermite quadratures as was done in \cite{kadoh2019}. We choose another way, which we feel is more transparent and gives a better control of the error.

It is not possible to carry out the field integrals independently of each other because of the $\exp(\phi_\ell\,\phi_r)$ term  coming from the expansion of the kinetic term in $F$. We can simply solve this problem by expanding the exponential in Taylor series 
\begin{equation}
    \exp(\phi_\ell\,\phi_r) = \sum_{n=0}^{+\infty} \frac{\phi_\ell^n \phi_r^n}{n!} \, ,
\end{equation}
which allows to carry out all the integrals independently. This yields a factorization of $F$ that can be pictorially represented as 
\begin{equation*}
    \matrixInit{1} = \matrixInit{4}  = \sum_{n=0}^{+\infty} \matrixInit{2} ,
\end{equation*}
where
\begin{equation}
    \matrixInit{3} \equiv \frac{\phi^n}{\sqrt{n!}}\exp \! \bigg[- \Big(\frac{1}{2}+\frac{\mu^2_0}{8}\Big)\phi^2 - \frac{\lambda}{16}\phi^4\bigg]\, .
\end{equation}
This effectively replaces integrals by discrete (infinite) sums. Let us perform this factorization for every $F$ in the tensor network:
\begin{equation*}
    \networkOG{1} \to \networkOG{2} \, ,
\end{equation*}
where we only represent a patch of the network, leaving the periodic boundary conditions implicit. Let us now define the elementary tensor $ T_{abcd}$ with discrete indices:
\begin{align}
    \label{eq:defT}
    T_{abcd}&\equiv\elementTensor{1} 
    =
    \elementTensor{2}
    \\
    \nn & =\int_\mathbb{R} \upd \phi \; \frac{\phi^{a+b+c+d}}{\sqrt{a!\, b!\, c!\, d!}} \, \exp \! \bigg[ \Big( \! -\frac{\mu_0^2}{2}- 2\Big) \, \phi^2  - \frac{\lambda}{4}\phi^4 \bigg] 
    \\[0.5em]
    & =\frac{h\left(\frac{a+b+c+d}{2},\mu_0,\lambda\right)}{\sqrt{a!\, b!\, c!\, d!}} \nn
\end{align}
so that the previous tensor network expression reads
\begin{equation}
    \TRG{0} \, .
\end{equation}
For a $2 \times 2$ square lattice, the partition function can now be depicted as
\begin{equation}
    \mathcal{Z}^{\mathsf a}_{2 \times 2 } = \partitionFunction{1} \, .
\end{equation}
The indices $(a,b,c,d) \in \mathbb{N}^4$ summed over still need to be restricted to a finite range $\llbracket K_\text{min},K_\text{max}\rrbracket $ for numerics. The advantage of this expansion is that the truncation error is precisely controllable and decreases faster than exponentially in the cut-off $K_\text{max}$ (see app.~\ref{appendix:discretization}). Finally, the coefficients $T_{abcd}$ can be easily evaluated in terms of special functions, namely for $n \in \mathbb{N}$
\begin{align}
    &h(n,\mu_0,\lambda) :=
    \!\! \int_\mathbb{R} \! \upd \phi\,\phi^{2n}
    \exp \!\! \bigg[ \Big( \! -\frac{\mu_0^2}{2}- 2 \Big) \phi^2  - \frac{\lambda}{4}\phi^4 \bigg] \\
    & \q =
    \Big(\frac{1}{\lambda}\Big)^{\frac{n}{2}+\frac{1}{4}}
    \Gamma\Big[n+\frac{1}{2}\Big]\, 
    U\bigg(\frac{1}{4}+\frac{n}{2}, \frac{1}{2}, \Big[2+\frac{\mu_0^2}{2}\Big]^2 \frac{1}{\lambda}\bigg) \, ,
    \nn
\end{align}
where the confluent hypergeometric function of the second kind $U(\alpha,\beta,z)$ is defined as
\begin{equation}
    U(\alpha,\beta,z)=\frac{1}{\Gamma(\alpha)} \int_0^{+\infty} 
    \!\!\! \upd t \; \e^{-zt} t^{\alpha-1}(1+t)^{\beta-\alpha-1} \, . 
\end{equation}
The latter has efficient implementations in most programming languages, and in particular in \texttt{mpmath} \cite{mpmath}. 

For half integers, $h$ vanishes, and thus $T_{abcd}\neq 0$ only if $a+b+c+d$ is even, which is a consequence of the $\mathbb{Z}_2$ symmetry of the problem. We exploit this feature in our algorithm by defining $\mathbb Z_2$ symmetry preserving tensors as described in \cite{Singh_2010,Singh_2011}, which significantly reduces the computational time.

\section{Renormalization in the tensor network language\label{sec:Gilt}}

\noindent
\emph{In this section, we explain how the renormalization group flow of $\phi^4_{2}$ can be computed in the context of tensor networks using the \textsf{Gilt}-\textsf{TNR} algorithm. For a more general and detailed description of this algorithm, we encourage the reader to consult \cite{hauru2018}}.

\subsection{Heuristics}

\noindent
We explained in the previous section how the partition function of the theory can be decomposed as a \emph{homogeneous} tensor network, whose underlying graph is a square lattice with periodic boundary conditions. In this context, computing the partition function boils down to contracting the corresponding tensor network. In order to obtain an accurate prediction of the partition function in the thermodynamic limit, the network should be taken to be as large as possible.

Since we are dealing with a homogeneous network, it could be systematically contracted by first blocking a square of four tensors, and then iterate the process after replacing the original tensor with the result of the first blocking.
Given an initial bond dimension $\chi$ along all the legs, one such round of contractions would result in a new tensor with bond dimension $\chi^2$, \ie
\begin{equation*}
    \dummyContraction{0}
    \to
    \dummyContraction{1}
    \to
    \dummyContraction{2}
    \, .
\end{equation*}
The cost of such an operation would therefore grow \emph{exponentially} in the lattice size, meaning only relatively small networks could be contracted in practice.

In order to deal with larger systems, some approximations need to be made, in such a way that the computational time of the contraction scheme only grows polynomially in the lattice size. If we are merely interested in the actual value of the partition function, then many options are available. However, our focus is rather on the renormalization group of the theory. This means that we want a method allowing us to define a flow in the space of tensors, obtained by defining effective tensors representing the same system at different length scales. 

In order to obtain an effective tensor that represent the infrared length scale of the theory, the bond dimension must be prevented from growing at every iteration of the blocking scheme. In general, this is done by replacing a local patch of tensors by another one, for which the bond dimensions along some of the legs have been reduced. 
Naturally, such a replacement must only induce a small error, otherwise the benefit of dealing with a larger lattice size would not outweigh the penalty of performing approximations.  This is guaranteed as long as some elements of the tensor account for \emph{short-range} physics only, and can therefore be safely discarded as they become irrelevant at larger scales. 

\subsection{Coarse-graining}

\noindent
The first algorithm to follow the philosophy described above was the so-called Tensor Renormalization Group (\textsf{TRG}) algorithm \cite{Levin:2006jai}. This algorithm relies on replacing single tensors by their \emph{truncated singular value decomposition}. Given a $p \times q$ matrix $M$, its singular value decomposition (SVD) is a factorization of the form $M = USV^\dagger$, where $U^\dagger U = \mathbbm 1_{p \times p }$, $V ^\dagger V = \mathbbm 1_{q \times q}$ and $S$ is $p \times q$ diagonal matrix. The entries $S_a := S_{ab}\delta_{ab} \geq 0$ are known as the \emph{singular values} of $M$. A \emph{truncated} singular value decomposition is then obtained by discarding some of the smallest singular values in the factorization above, \ie it is a low-rank approximation of the form
\begin{equation}
    M_{ab} \approx \sum_{k=1}^\chi U_{ak}S_k V_{bk}^* \, ,
\end{equation}
where $\chi < {\rm rank}(M)$. This definition is easily generalized to tensors and can be schematically illustrated as
\begin{equation*}
    \SVD{1}    
    \; \stackrel{{\rm svd}}{=} \; 
    \SVD{2}
    \; = \;
    \SVD{3} \, ,
\end{equation*}
where the dotted lines indicate that there is an arbitrary number of legs on both side of the tensor $A$.
By blocking $\{U,S^\frac{1}{2}\}$ on the left-hand side, and $\{S^\frac{1}{2},V\}$ on the right-hand side, we obtain a \emph{split} of the tensor:
\begin{equation*}
    \raisebox{2pt}{\SVD{1}}    
     \stackrel{{\rm svd}}{=} \;
    \SVD{8}  \; ,
\end{equation*}
where 
\begin{equation*}
    \SVD{5} \; := \; \SVD{4} \q {\rm and} \q \SVD{7} \; := \; \SVD{6} \; .
\end{equation*}
Replacing the SVD by a truncated SVD in this process yields an \emph{approximate split} of the tensor. The key ingredient of the \textsf{TRG} algorithm is the replacement of all the tensors by their approximate splits. More precisely, for a square lattice, the algorithm proceeds in two steps: First, the initial tensors are split via truncated singular value decompositions along two orthogonal directions:
\begin{equation}
    \label{eq:TRGflow1}
    \TRG{1} \xrightarrow{{\rm svd}} \TRG{2} \, .
\end{equation}
Second, four tensors resulting from the splittings are contracted together to define a new four-valent tensor: 
\begin{equation*}
    \begin{tikzpicture}[scale=0.85,baseline={([yshift=-.5ex]current bounding box.center)}]
        \draw[] (-0.5,-0.5) -- (0.5,0.5);
        \draw[] (-0.5,0.5) -- (0.5,-0.5);
        \tenBig{0,0}{ten0};
    \end{tikzpicture}
    \equiv
    \begin{tikzpicture}[scale=0.85,baseline={([yshift=-.5ex]current bounding box.center)}]
        \TRGUnitCell{0}{0}
    \end{tikzpicture} \; .
\end{equation*}
This results in a new homogeneous tensor network whose underlying graph is still a square lattice, but rotated by $45^\circ$: 
\begin{equation}
    \label{eq:TRGflow2}
    \TRG{3} \, .
\end{equation}
Repeating the same two steps would yield a square lattice in the original orientation, with an effective lattice spacing twice as large as the original one. Since the truncated SVDs prevent the bond dimensions from growing uncontrollably high, this \emph{coarse-graining} scheme can be iterated, hence yielding a flow in the space of tensors from UV to IR.

\subsection{Conceptual shortcomings}
\noindent
The \textsf{TRG} coarse-graining scheme is undoubtedly simple and has proven very successful to study two-dimensional classical systems and ground or thermal states of one-dimensional quantum Hamiltonians.\footnote{Given a one-dimensional quantum Hamiltonian $\mathbb H$, the Euclidean path integral $e^{-\beta \mathbb H}$ can be rewritten as a tensor network, whose underlying graph is also a square lattice, via a Suzuki-Trotter decomposition \cite{SUZUKI1990319, 1991JMP....32..400S}.} Despite its success, the conceptual shortcomings of this simple approach have been known for several years. Indeed, it was emphasized in \cite{Evenbly_2015} that, in spite of its name, \textsf{TRG} does not yield a \emph{proper} RG flow.

This is seen easily by inspecting the fixed point structure of the algorithm. Indeed, a simple analysis concludes that the fixed points of the flow generated by \textsf{TRG} depend on the initial tensor, meaning that they display features that are \emph{non-universal}. This goes against the spirit of RG which states that upon coarse-graining, the UV details below the scale considered should be integrated away. Translating this statement into our language, this signifies that all information contained within the tensors describing short-range physics at a given scale should be contracted during a coarse-graining step. But, perhaps counter-intuitively, the \textsf{TRG} algorithm contracts away only a fraction of short-range information. 

This is typically illustrated by means of a toy model for purely short-range physics known as the corner-double-line (CDL) tensor. Given an arbitrary matrix $M$, a four-valent CDL tensor is of the form
\begin{align*}
    T^{\rm CDL} \equiv \CDLTensor \;\;\; \text{with} \;\;\;
    T^{\rm CDL}_{abcd}& \equiv T^{\rm CDL}_{(a_1a_2)(b_1b_2)(c_1c_2)(d_1d_2)}
    \\
    :\!\!&= M_{a_1b_2}M_{b_1c_2}M_{c_1d_2}M_{d_1a_2} \,.
\end{align*}
When organized in a square lattice, these CDL tensors give rise to so-called \emph{loops of short-range correlations} within each square plaquette. These loops indicate that any observable supported on a given leg of the network would be highly correlated with any other observable that has support around one of the plaquettes that include this leg. Conversely, these loops are completely uncorrelated with anything beyond their immediate surrounding. Upon a coarse-graining step of the \textsf{TRG} algorithm, half of the loops are traced away when blocking the tensors resulting from the splits. However, the other half remain untouched and become purely short-range information with respect to the new length scale:
\begin{equation*}
    \TRG{4} \to \TRG{5} \, .
\end{equation*}
Consequently, although these CDL tensors are only non-trivial at the length scale of the lattice, they form fixed points of the flow yielded by \textsf{TRG}.

More generally, all the short-range physics details that are not removed at a given coarse-graining step of \textsf{TRG} accumulate, polluting the coarse-grained tensor and therefore the RG flow. As we already mentioned, this implies that the fixed point structure of the flow is not physically correct, making the identification of the genuine IR fixed points more difficult. Quantitatively, flowing irrelevant UV information in the tensor at each step reduces the space available to store the relevant IR information, and thus the accuracy of predictions at fixed bond dimension.

Crucially, these difficulties become particularly pregnant when studying critical points, making the scale invariance near impossible to witness, and the measurement of critical exponents less reliable \cite{hauru2018}. Note that by allowing for higher bond dimensions, the quantitative predictions of \textsf{TRG} can be naturally improved. However, because of its failure to contract away all the UV details at every step, it cannot yield a proper RG flow no matter the bond dimension. In many cases, these conceptual shortcomings do not drastically undermine the usefulness of this algorithm, but if we are interested in the RG flow, then it is crucial to use an algorithm that addresses them explicitly.

\subsection{Graph independent local truncations}
\noindent
In order to remove the short-range information that survive the \textsf{TRG} coarse-graining, more truncations need to be made by means of local replacements. But these additional local replacements should be such that the geometry and homogeneity of the network are preserved at every step, so that a flow in the space of tensors can still be defined. Over the years, several schemes that address the shortcomings of \textsf{TRG} and yield proper RG flows have been constructed, \eg Tensor Network Renormalization \textsf{TNR} \cite{Evenbly_2015}, Loop-\textsf{TNR} \cite{Yang_2017} or \textsf{TNR+} \cite{Bal_2017}. In this manuscript, we use a more recent approach based on an algorithm referred to as \textsf{Gilt}, which performs \emph{graph independent local truncations} \cite{hauru2018}. 

\textsf{Gilt} is a general-purpose algorithm that performs local truncations of bond dimensions in any tensor network without affecting its geometry. We use it to truncate the bond dimensions of carefully chosen individual legs of the network, thereby removing the short-range information present in some environment surrounding them. 

Let us consider a plaquette of the network and pick a leg that is included in this plaquette. Let $\chi$ be the bond dimension of this leg. The \textsf{Gilt} algorithm consists in replacing the leg, which can be thought as being the identity matrix, by a non-trivial matrix. This non-trivial matrix is designed to be as low-rank as possible. That way, by splitting the matrix and contracting the components of this split with the neighbouring tensors, we effectively truncate the bond dimension of the original leg. This process can be summarized as follows:
\begin{equation*}
    \GiltOpt{1} 
    \approx
    \GiltOpt{2}
    \stackrel{\rm svd}{=}
    \GiltOpt{3}
    =
    \GiltOpt{4} \, ,
\end{equation*}
with $\chi ' < \chi$. Let us now explain how to design such low-rank matrix. We shall merely provide the solution and refer the reader to \cite{hauru2018} for additional details and motivation. 

Let us consider the network obtained by removing the leg whose bond dimension we wish to truncate, and perform the SVD illustrated below:
\begin{equation}
    \label{eq:GiltSVD}
    \GiltOptSVD{1} \equiv \GiltOptSVD{2} \stackrel{\rm svd}{=}
    \GiltOptSVD{3} \; .
\end{equation}
The first identification indicates that we consider the blocking of the four tensors as a single tensor. We then perform the SVD of the matrix obtained by grouping the two middle upper most legs together, providing one index of the matrix, and all the other legs together, providing the other index. Notice that we do not truncate any bond dimension at this point. The tensor $U$ provided by the SVD may be used to form a resolution of the identity, which we insert on the original leg of the network
\begin{equation}
    \label{eq:GiltOpt}
    \GiltOpt{101} 
    = 
    \GiltOpt{5}
    \equiv
    \GiltOpt{6}\, .
\end{equation}
In the equation above, we introduced a special notation for the partial trace of $U$ such that $t_i = {\rm tr} \, U_i$. 
We are now ready to define the \textsf{Gilt}-matrix as
\begin{equation}
    \GiltMatrix{1} := \GiltMatrix{2} \, ,
\end{equation}
where the vector $t'$ is defined in terms of $t$, the singular values of the SVD \eqref{eq:GiltSVD}, and a parameter $\epsilon_\textsf{Gilt}$ as 
\begin{equation}
    \label{eq:GiltMatrix}
    t'_i := t_i \, \frac{S_i^2}{S_i^2 + \epsilon_\textsf{Gilt}^2} \, ,
\end{equation}
which effectively provides a soft truncation of the singular values of $S$.
This solution was found in \cite{hauru2018} to be very good, although not necessarily optimal, to minimize the rank of the \textsf{Gilt}-matrix while maximizing the overlap between the original network and the one after inserting the \textsf{Gilt}-matrix. The hyper-parameter $\epsilon_\textsf{Gilt}$ is chosen by the user depending on the variables of the model and the parameters of the simulation, \eg the maximal bond dimension authorized.

Intuitively, $\epsilon_\textsf{Gilt}$ controls the trade-off between rank reduction and global accuracy. A large $\epsilon_\textsf{Gilt}$ is likely to substantially reduce the rank of the \textsf{Gilt}-matrix and remove more irrelevant short-range information. However, it might introduce a larger local error.  Conversely, a small $\epsilon_\textsf{Gilt}$ is less likely to reduce the rank of the \textsf{Gilt}-matrix, leaving us in a situation akin to that of \textsf{TRG}, where irrelevant information propagates from a scale to the next, ultimately taking up all the parameters of our finite $\chi$ tensor. In app.~\ref{sec:app_chiAndEps}, we explain our procedure to choose $\epsilon_\mathsf{Gilt}$.

\subsection{Renormalization scheme}
\noindent
Recall that the failure of \textsf{TRG} to yield a proper RG flow stems from its inability to remove the short-range information within half of the plaquettes. It was shown in \cite{hauru2018} that by applying the \textsf{Gilt} algorithm along the edges surrounding the plaquettes at issue, the short-range information that survive the coarse-graining can be truncated away. More specifically, given such a plaquette, we iterate the \textsf{Gilt} algorithm on all the legs surrounding it, effectively removing the loop of correlations contained within:\footnote{In order to significantly reduce the bond dimension, it is often necessary to iterate the \textsf{Gilt} algorithm on a single leg before moving on to another one. However, this converges quickly and the limiting step, which is the initial SVD, only needs to be performed once.}
\begin{equation*}
    \miniTNR{0} \approx \miniTNR{1} \stackrel{\rm svd}{=} \miniTNR{2} = \miniTNR{3}.
\end{equation*}
Notice that the \textsf{Gilt} matrices generally differ from one leg to another. Nevertheless, since the network is homogeneous and we only need to apply the \textsf{Gilt} algorithm on every other plaquette, only four matrices need to be computed throughout.

\begin{figure*}[t]
    \centering
    \includegraphics[scale=1]{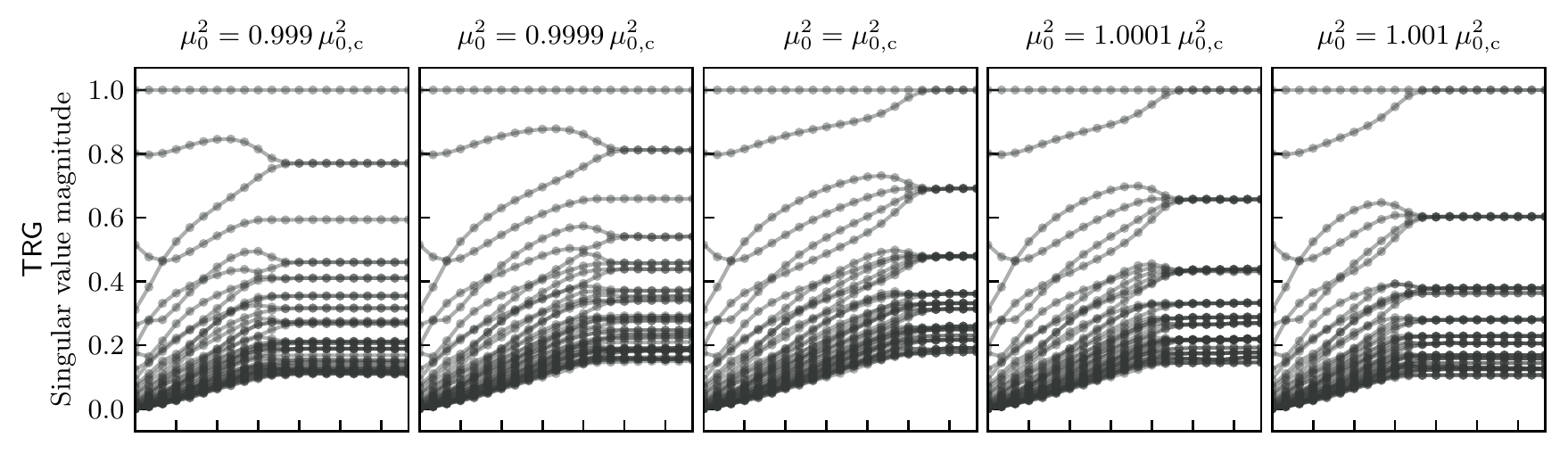}
    \includegraphics[scale=1]{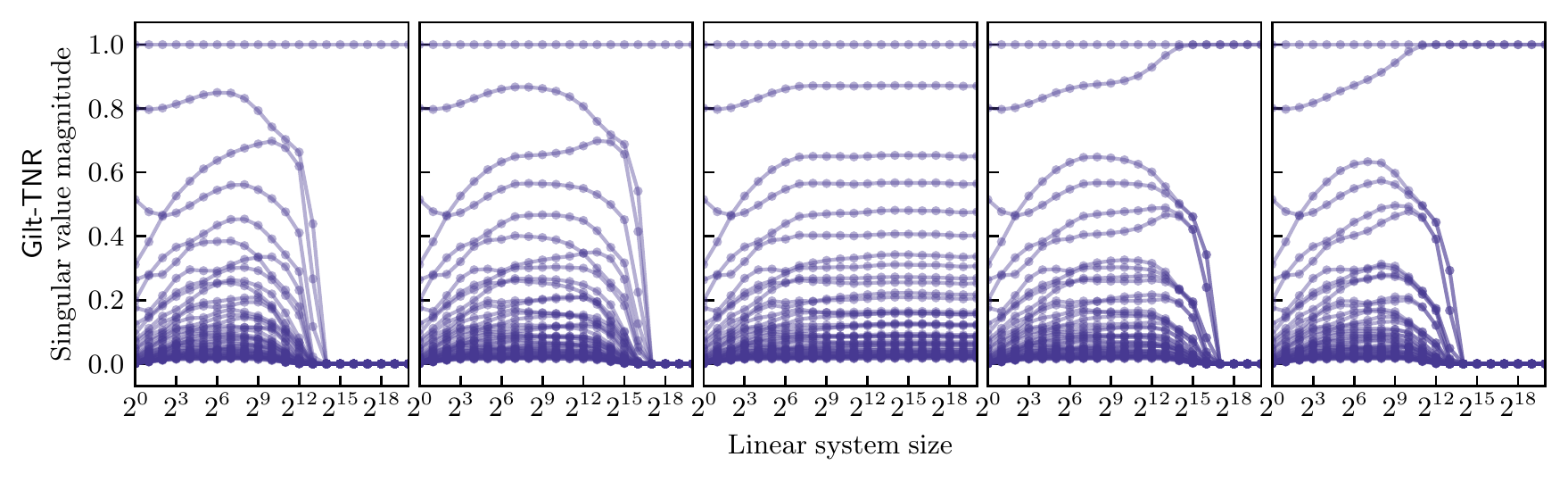}
    \caption{RG flow in the space of tensors for $\lambda=0.1$ at five different values of $\mu_0^2$ using \textsf{TRG} (upper panel) and \textsf{Gilt-TNR} (lower panel). For every linear system size (or equivalently number of RG steps), the 45 largest singular values of the coarse-grained tensors computed according to \eqref{eq:spectrum_SVD} are displayed. Each line corresponds to the flow of a given singular value. For both \textsf{TRG} and \textsf{Gilt-TNR}, the maximal bond dimension is $\chi=110$, and the threshold parameter was chosen to be $\epsilon_\textsf{Gilt}=5 \cdot 10^{-8}$ for \textsf{Gilt-TNR}. Whereas \textsf{TRG} yields non-universal fixed points, \textsf{Gilt-TNR} provides the right fixed point structure away from criticality and reveals a non-trivial fixed point at criticality that is characteristic of the underlying CFT.}
    \label{fig:flow_spectra}
\end{figure*}

By preceding every iteration of \textsf{TRG} with an iteration of \textsf{Gilt}, we obtain the renormalization scheme referred to \textsf{Gilt}-\textsf{TNR}. First, \textsf{Gilt} matrices are inserted on  the legs surrounding every other plaquette, which are then split via truncated SVDs:
\begin{equation*}
    \TNR{1} \xrightarrow{{\rm svd}} \TNR{2} \, .
\end{equation*}
Blocking four components of the splits around every tensor of the original network, we obtain two different tensors:
\begin{equation*}
    \GiltTensor{3}
    \equiv
    \GiltTensor{2}
    \q {\rm and} \q
    \GiltTensor{1}
    \equiv
    \GiltTensor{4} \, .
\end{equation*}
At this point, we are in the initial configuration of \eqref{eq:TRGflow1} with the difference that we distinguish two types of tensors and half the loop of correlations have been removed thanks to \textsf{Gilt}. \textsf{TRG} can now be applied
\begin{equation*}
    \TNR{3} \xrightarrow{{\rm svd}} \TNR{4} \, .
\end{equation*}
It remains to block sets of four tensors as follows
\begin{equation*}
    \begin{tikzpicture}[scale=0.85,baseline={([yshift=-.5ex]current bounding box.center)}]
        \draw[] (-0.5,-0.5) -- (0.5,0.5);
        \draw[] (-0.5,0.5) -- (0.5,-0.5);
        \tenBig{0,0}{ten0};
    \end{tikzpicture}
    \equiv
    \begin{tikzpicture}[scale=0.85,baseline={([yshift=-.5ex]current bounding box.center)}]
        \TRGUnitCellBi{0}{0}{ten2}{ten5}
    \end{tikzpicture} \, ,
\end{equation*}
and in the process remove the last loops of short-range information. The resulting network is still homogeneous so that the whole algorithm can be iterated until the desired infrared length scale is reached.

\subsection{Extracting observables\label{sec:extract}}

\noindent
As mentioned earlier, we are interested in the RG flow in the space of tensors generated by the \textsf{Gilt-TNR} algorithm. Concretely, the flow is analysed by computing the \emph{spectrum} of the tensor after every iteration of the algorithm, where by spectrum we mean the largest singular values of the following decomposition:
\begin{equation}
    \label{eq:spectrum_SVD}
    \spectrum{1} \stackrel{\rm svd}{=} \spectrum{2}\, .
\end{equation}
This spectrum provides a precise and basis independent characterization of the tensor. In particular, it allows for the immediate identification of the trivial fixed points of the RG flow of $\phi^4_{2}$, characterized by the number of non-vanishing degenerate singular values. Once we have identified the trivial fixed points, we can proceed by dichotomy to find the critical points at which phase transitions take place. This method is simple, robust and its accuracy only depends on the resolution of the input parameters.

As a benchmark, it is also useful to compute the scaling dimensions of the conformal theory at criticality. These observables are also straightforwardly obtained by diagonalizing the \emph{transfer matrix} resulting from the contraction of two coarse-grained tensors as follows:
\begin{equation}
    \mathbb T := \transfer \, ,
\end{equation}
which is related via a logarithmic conformal map to a discrete version of the dilation operator \cite{Evenbly_2016}.

\begin{figure*}[t]
    \hspace{-0.5em}\includegraphics[scale=1]{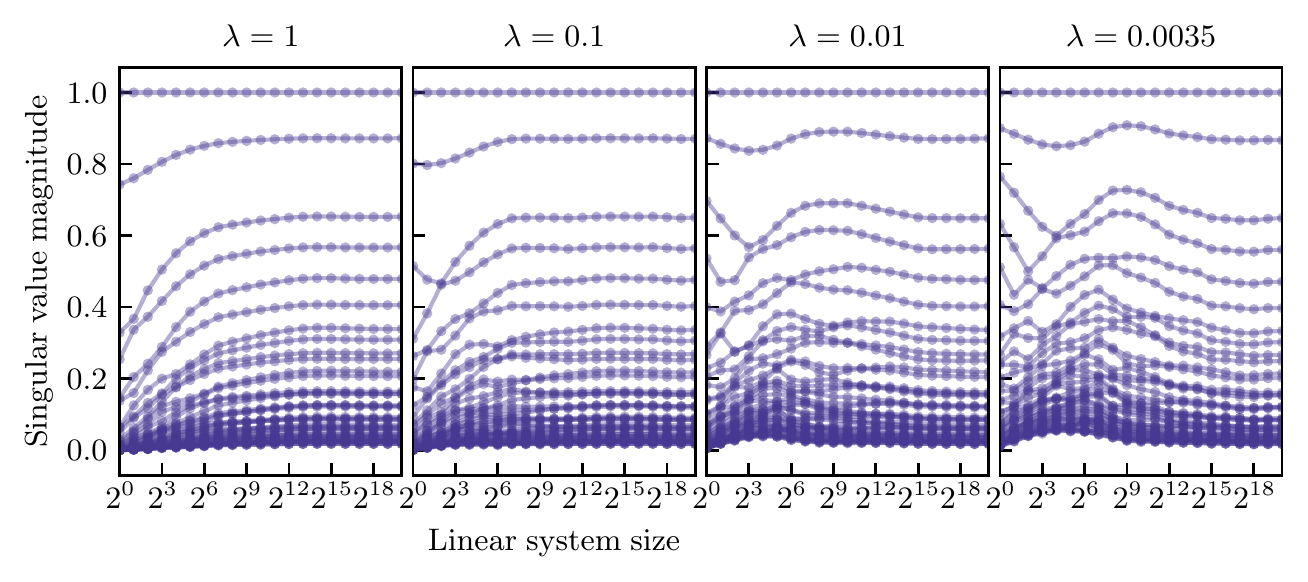} \hspace{-1.83em}
    \heuristicFlow
    \caption{RG flows in the space of tensors at criticality for four different values of $\lambda$ using the \textsf{Gilt-TNR} algorithm at bond dimension $\chi = 110$. After a number of iterations that grows as $\lambda$ gets close to zero, the same non-trivial fixed point, which is characteristic of the underlying CFT, is observed. On the right-hand panel is depicted a heuristic picture of the RG flow, which illustrates the fact that as we go to the continuum limit, the QFT UV fixed point, with a continuum of eigenvalues, starts being visible.}
    \label{fig:heuristicFlow}
\end{figure*}

\section{Results\label{sec:results}}

\noindent
\emph{In this section we present the results of applying the \textsf{Gilt-TNR} algorithm to $\phi^4_{2}$}.

\subsection{Renormalization group flow}

\noindent
Let us first analyse the RG flow of the lattice theory for a given $\lambda$ that is not too small. In fig.~\ref{fig:flow_spectra}, we present the RG flows generated by the \textsf{Gilt-TNR} and \textsf{TRG} algorithms for $\lambda=0.1$. We display on these graphs the spectra as defined in the previous section for coarse-grained tensors corresponding to different linear system sizes. Five pairs of panels are presented corresponding to five different values of $\mu_{0}^2$: the one at criticality, and two on both sides of the critical point. 

We notice immediately the difference in fixed point structures. First of all, the \textsf{Gilt-TNR} algorithm yields the right trivial fixed points regardless of the initial parameters. These are characterized by a single or a pair of dominant singular values, that correspond to either the disordered phase or the symmetry broken phase, respectively. The \textsf{TRG} algorithm yields non-universal fixed points that depend on the initial parameters. While the non-universal features that correspond to IR-irrelevant information are seen to accumulate as we get closer to the critical point, they are systematically removed by \textsf{Gilt}, eventually leading to a collapse of the singular values, producing the trivial fixed points.  

Strikingly, \textsf{Gilt-TNR} yields a non-trivial fixed point at critically that is characteristic of the underlying CFT. Indeed, we observe that as the system size grows, the spectrum remains approximately identical for many iterations, which is  the hallmark of scale invariance. This cannot be observed using \textsf{TRG} because of the accumulation of short-range information that  requires an unmanageable growth of the bond dimension.

When going to the continuum limit by sending $\lambda$ to zero, the RG picture becomes more interesting. We naturally have the same IR phases and fixed points, but a new UV fixed point -- corresponding to the short distance regime of the continuum QFT -- appears. Indeed, we now start from an ``ultra-UV'' lattice theory, then flow close to the QFT UV fixed point (the massless boson) as irrelevant lattice artifacts decay, and then only reach the IR regime of the QFT, which is independent of the continuum limit. The lattice artifacts at the start of the RG are the same no matter the initial value of $\lambda$, but only with very small values of $\lambda$ can the system flow close to the UV QFT before the relevant perturbations (mass and coupling) drive it away to the infrared (see fig.~\ref{fig:heuristicFlow}). The traces of the UV QFT are clear in the tensor spectrum. As $\lambda$ goes to zero, we observe that the initial singular values are systematically larger. We then observe a decay of the first singular values that is more pronounced as $\lambda$ is small, sign that we are near a fixed point with a continuum of eigenvalues, which is the case for the free boson. As we argue in sec.~\ref{sec:discussion}, this makes going to the continuum limit expensive.

\subsection{Critical coupling in the continuum}

\begin{figure*}[t]
    \centering
    \includegraphics[scale=.89]{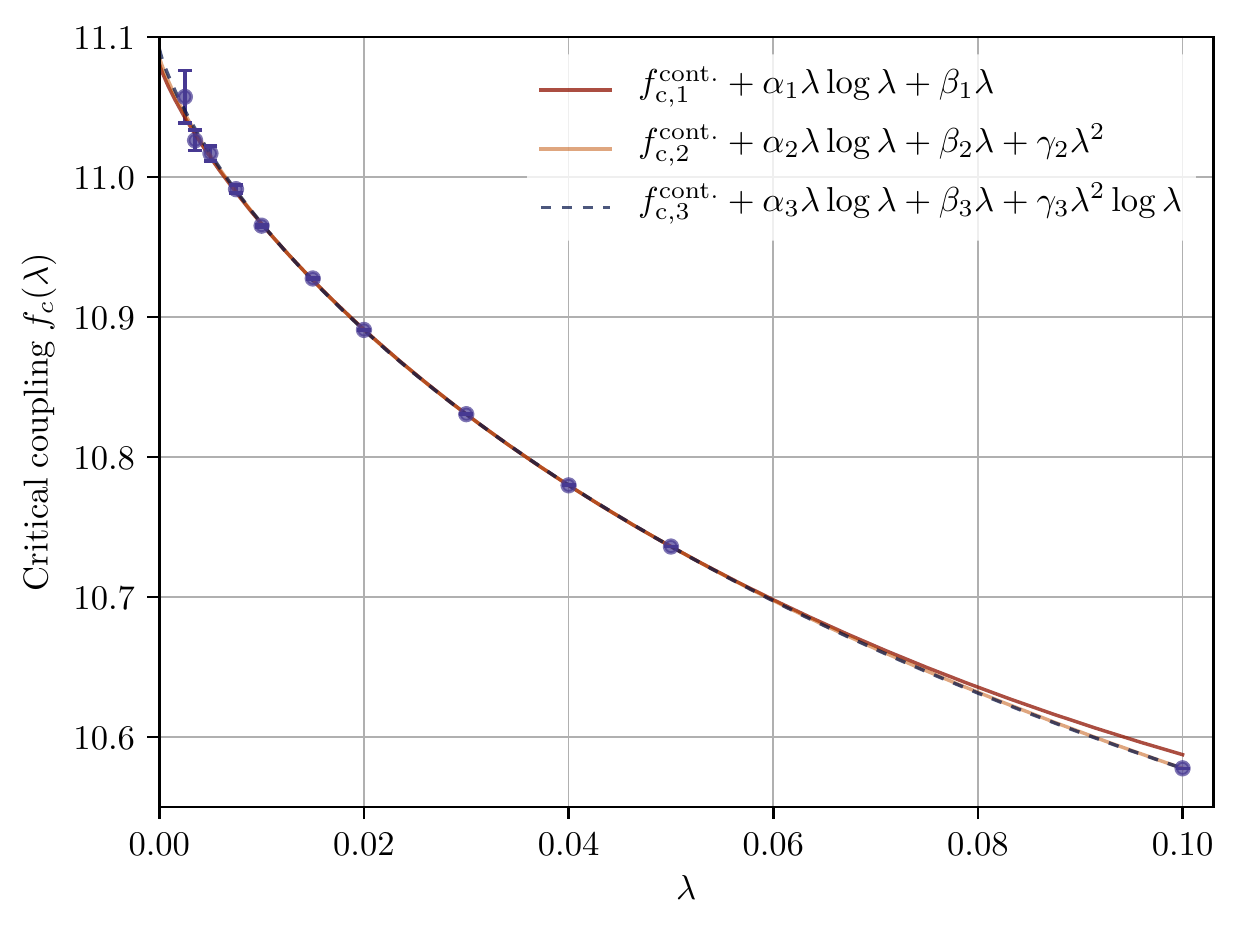}
        \includegraphics[scale=.89]{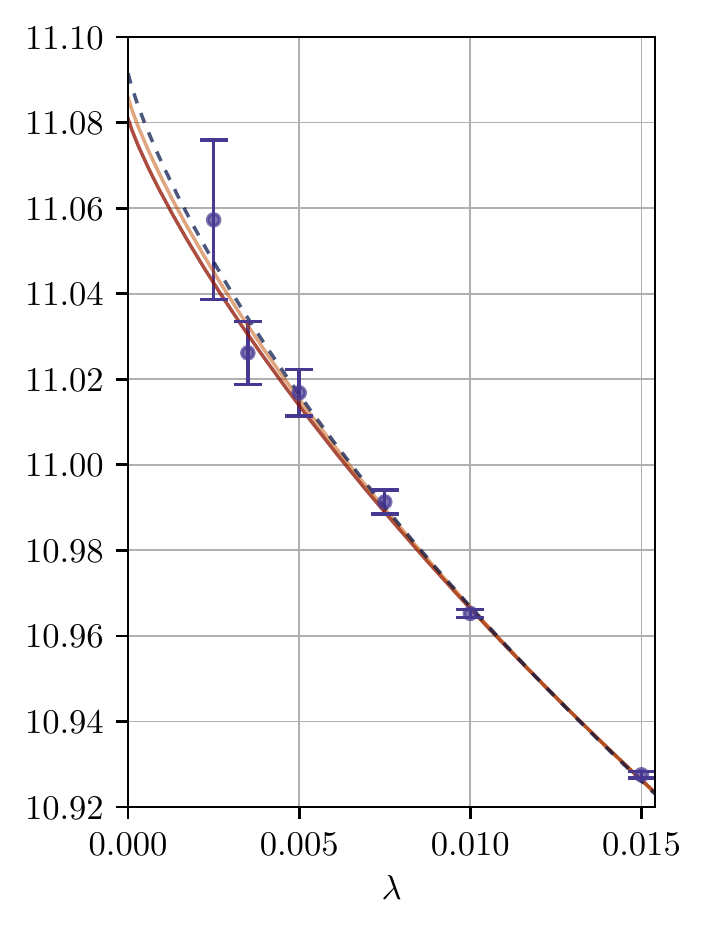}
    \caption{Critical coupling $f_c=\lambda/{\mu_{\rm c}^2}$ as a function of $\lambda$, computed for $\chi=110$ and $\epsilon_\mathsf{Gilt}=5 \cdot 10^{-8}$. The error bars are estimated as described in app.~\ref{sec:app_chiAndEps}. The first fit up to linear order does not include the point $\lambda=0.1$ as it is expected to be valid only for smaller values of the coupling. The right-hand panel shows a narrower view of the same figure close to $\lambda=0$. }
    \label{fig:fc}
\end{figure*}

\begin{table}[b]
    \setlength{\tabcolsep}{.48em}
    \begin{tabularx}{\columnwidth}{lllc}
        Method & $f_{\rm c}^\text{cont.}$  & Year & Ref. \\
        \midrule
        Tensor network coarse-graining  & 10.913(56) & 2019 & \cite{kadoh2019}\\
        Borel resummation & 11.23(14) & 2018 & \cite{serone2018}  \\%
        Renormalized Hamil. Trunc. & 11.04(12) & 2017 & \cite{eliasmiro2017-2}\\
        Matrix Product States  & 11.064(20) & 2013 & \cite{milsted2013}\\
        Monte Carlo   & 11.055(20) & 2019 & \cite{bronzin2019} \\
        This work  & \fc  & 2020 & {}
    \end{tabularx}
    \caption{%
        Comparison of several estimates of the critical coupling constant $f_{\rm c}^\text{cont.}$ in the continuum obtained using different methods. 
    }
    \label{tab:summary_f}
\end{table}

\noindent
Analysing the renormalization group flow of the model as described above, we can determine the critical coupling $f_{\rm c}(\lambda)$ of the lattice theory for any values of $\lambda$. Since we are interested in the continuum field theory, we probe small values of $\lambda$, going as far as $\lambda=0.0025$. We then extrapolate to the continuum limit. The values of $f_{\rm c}(\lambda)$ for finite $\lambda$ are shown in fig.~\ref{fig:fc}. 

The precision of our data points leaves little doubt that the leading term in the fit to the continuum limit is not linear but $\propto \lambda \log \lambda$. More precisely, we fit the three following functions to extrapolate to the continuum limit:
\begin{align}
\label{eq:fit}
    &f_{{\rm c},1}^\text{cont.} \! + \alpha_1 \, \lambda \log \lambda + \beta_1 \,\lambda \, ,&  \upchi^2/\text{d.o.f.}= 0.96 \, , \nn \\
    &f_{{\rm c},2}^\text{cont.} \! + \alpha_2 \, \lambda \log \lambda + \beta_2 \,\lambda + \gamma_2\, \lambda^2 \, ,& \upchi^2/\text{d.o.f.}= 1.14 \, , \nn \\
    &f_{{\rm c},3}^\text{cont.} \! + \alpha_3 \, \lambda \log \lambda + \beta_3 \,\lambda + \gamma_3 \lambda^2\log\lambda \, ,&\upchi^2/\text{d.o.f.}= 1.53 \, ,
    \nn
\end{align}
where $\upchi^2/\text{d.o.f.}$ is the ``variance of unit weight'' of the weighted least square fit\footnote{A value of $\upchi^2/\text{d.o.f.} \gg 1$ means that the realized variance is larger than if the model were correct, which suggests we either lack important terms in the fit or that we underestimate the error in our data points. On the other hand, $\upchi^2/\text{d.o.f.} \ll 1$ means that we are either over-fitting the data or that the error bars we report are overestimated. Values close to $1$ are compatible with an appropriate expansion and well-estimated errors. Note that for a quadratic fit without $\log$ terms, for $\lambda \in [0.0025,0.05]$, we find $\upchi^2/\text{d.o.f.} = 10.4$.}.
This gives
\begin{align}
    \nn
    f^{\rm cont.}_{{\rm c},1} &= 11.0809(15)\, ,\\
    f^{\rm cont.}_{{\rm c},2} &= 11.0859(22) \, ,\\
    \nn
    f^{\rm cont.}_{{\rm c},3} &= 11.0915(32) \, .
\end{align}
A rigorous analysis of the errors of our algorithm is difficult. Indeed, although we know the errors associated with every approximate replacement performed throughout the algorithm, we are unaware of any methods to accurately estimate the global error from them. As we explain in app.~\ref{sec:app_chiAndEps}, we crudely estimate the error bars from the fluctuations of our results as a function of the bond dimension. We notice that the uncertainty associated with the choice of fit is greater than the uncertainty within each fit. To be as conservative as possible, we average the final result over the three fits and quote as error the distance between the two extreme fits majored of their associated errors. We thus quote  $f_{\rm c}^\text{cont.}=\fc$, a value which favourably compares with previous studies (see table \ref{tab:summary_f}).

The existence of a $\lambda\log\lambda$ has been an open question in the existing literature, with the most recent studies unable to know for sure if a logarithmic term was needed with the precision available \cite{bronzin2019}. We do not know of a theoretical argument for this scaling for the particular case of $\phi^4_{2}$, but our results strongly suggest that it should be included. Note that the presence of a logarithmic term explains the slight underestimation of the continuum critical coupling reported in most previous studies that did not consider it.

\subsection{Scaling dimensions}

\noindent
To illustrate the efficiency of the algorithm, it is interesting to present benchmark results for the scaling dimensions at criticality, which are known from conformal field theory. Recall that on the whole $(\lambda, \mu_{\rm c}(\lambda))$-line, the critical point is in the Ising universality class, and that the Ising model corresponds to $\lambda \to \infty$. 

\noindent
We report in table \ref{tab:summary_scaling}, the values we obtain for the first generations of scaling dimensions using the transfer matrix method reviewed briefly in sec.~\ref{sec:extract}. We do so for different values of $\lambda$ to showcase the universality of the critical line. Note that the best accuracy is usually obtained for the coarse-grained tensors corresponding to the length scale where scale invariance starts being visible when inspecting the spectrum aforementioned, \eg for a linear system size of $2^{11}$ for $\lambda=0.1$. 

We notice that the precision does not depend strongly on the value of $\lambda$, at odds with what we see for the critical coupling, where errors increase as we get closer to the continuum limit. This is because the critical exponents are universal and thus insensitive to microscopic details. This is where the fact that we do a proper RG is fundamental: we reach the correct IR fixed point even as our UV errors increase.

\begin{table}[t]
	\setlength{\tabcolsep}{0.71em}
	\begin{tabularx}{\columnwidth}{lllll}
		Exact    & $\lambda = 1$ & $\lambda = 0.1$ & $\lambda=0.01$ & $\lambda=0.0035$  \\%
		\midrule
		0.125     & 0.1250004 & 0.1250005 & 0.125003 & 0.125004 
		\\
		1    & 1.00003 & 1.00003 &  1.00001   & 0.99992 
		\\
		1.125    & 1.12504 & 1.12507 & 1.12507 & 1.12507 
		\\
		1.125   & 1.12505 & 1.12507 & 1.12508  & 1.12509  
		\\
		2    & 2.0002 & 2.0003 & 2.0003 & 2.0002 
		\\
		2   & 2.0002 & 2.0003 & 2.0004 & 2.0004 
		\\
		2   & 2.0003 & 2.0005 & 2.0005 & 2.0005 
		\\
		2  & 2.0004 & 2.0005 & 2.0005 & 2.0006  
	\end{tabularx}
	\caption{%
		First scaling dimensions of the Ising CFT, as obtained by the transfer matrix method, for different values of $\lambda$.
	}
	\label{tab:summary_scaling}
\end{table}

\section{Discussion\label{sec:discussion}}

\noindent
\emph{We conclude this manuscript with a discussion of the errors and some comments regarding generalization to higher dimensions.}

\subsection{Understanding the sources of errors}

\noindent
We have shown  that using tensor networks can turn the renormalization group from a qualitative idea to a powerful computational tool. Nonetheless, if our results in the continuum limit are precise, they do no yet reach machine precision. This brings the question of what, if anything, limits the accuracy now and perhaps in a future with larger computing resources.

First, for a fixed spacing $\mathsf{a}$, tensor networks methods are extremely efficient. Although the evidence is not fully conclusive, it is believed that for the type of systems we consider, the \textsf{Gilt-TNR} algorithm makes errors that decrease exponentially in the maximum bond dimension $\chi$ \cite{hauru2018}. Other tensor network methods have a provable exponentially convergent error for local observables \cite{PhysRevLett.93.140402, Hastings_2007}, making the previous observation believable. The cost of the algorithm scales in $\mathcal{O}(\chi^7)$ in $d=2$ \cite{hauru2018}, which is stiff yet still polynomial. Hence, away from the continuum limit, the error can likely be arbitrarily reduced at a moderate cost (see app.~\ref{sec:app_chiAndEps} for brief discussion of empirical error estimation). 

Going to the continuum limit brings two additional sources of errors as compared to the setup with fixed lattice spacing: $(i)$ The bond dimension of the initial tensor grows linearly in the inverse lattice spacing $\mathsf{a}$ for a fixed error, $(ii)$ the RG has to be run for more steps as $\mathsf{a}\rightarrow 0$ to reach the same IR scale of the field theory. 

One could think that $(i)$ is a mere artifact of our discretization, which works well for small values of the field and strong coupling. But this is in fact independent of the basis chosen. Indeed, as we refine the lattice, the theory becomes free, and thus its spectrum becomes that of the free boson, which is continuous. Hence, as we get closer to the continuum limit, the singular values of the tensor decay slower and slower, and more and more of them have to be kept for a fixed error tolerance.

Difficulty $(ii)$ is less of a problem for our method because the scale we consider grows exponentially in the number of RG steps, and thus reducing the lattice spacing only incurs a logarithmic increase in the computational cost. Even for a very fine grained lattice, finite size effects (or IR cut-off effects) can thus be made negligible. This would be much less favorable for methods like the corner transfer matrix renormalization group (CTMRG) \cite{1996JPSJ...65..891N, 2009PhRvB..80i4403O, Orus:2011nj}, where the scale grows only linearly in the number of iterations.

For the smallest values of $\lambda$ we probed in this paper, we have reached the point where UV complications, related to driving the system towards the free boson continuum, start to outweigh the IR difficulties, related to the running of the RG flow near the Ising CFT fixed point. In particular, for $\lambda=0.0025$, the initial tensor is required to have a bond dimension that is significantly larger than the maximal one we allow in subsequent steps, in order to make sure that the field discretization error remains negligible.

While we do not believe there is any unacceptable bottleneck with current techniques, new methods to deal with the UV will likely need to be found to make the algorithm more efficient. One could try to use better discretizations of the gradient square term in the action so as to reduce lattice artifacts \cite{symanzik1983}. But the locality of the tensor network structure makes the introduction of next-nearest-neighbor terms less straightforward than in Monte-Carlo. It could also be useful to have a better theoretical control of the expansions of physical quantities away from the continuum as a function of $\lambda$. If these were well understood enough, in particular regarding peculiar logarithmic corrections, safer extrapolations to the continuum could be made without the need for extreme fine-graining.

Ironically, the regime we  struggle to capture is one that is trivial for perturbation theory, because it is essentially free. We speculate that there must be a way to use the best of the two approaches: Running the flow perturbatively for a while, until the theory is weakly coupled but not infinitely weakly coupled, and then run the hard strongly-coupled part of the flow with our method. We note that the recent bosonic-$\textsf{TRG}$ of Campos et al.~\cite{campos2019}, which uses continuous variables and Gaussian integrations to renormalize free field theories, seems that it could deal precisely with the regime that is hard to capture with our method. However, how to hybridize the two approaches is not obvious to us at the moment.

Let us finally mention that as we increase the number of physical dimensions to 3+1 and go to QFTs that are no longer super-renormalizable but just renormalizable, the previous UV difficulties will become milder. Indeed, even for an asymptotically free theory like quantum chromodynamics, the coupling constant is believed to run to zero only logarithmically in the lattice spacing (instead of polynomially in super-renormalizable ones). This means that we can start from a lattice spacing dozens of orders of magnitude smaller than the physically relevant length scales while keeping a theory that is not too weakly coupled and thus does not require a tensor with a pathologically large number of singular values. However, as we will see, increasing the number of dimensions brings other algorithmic difficulties.

\subsection{Going beyond two dimensions}

\noindent
In the future, the challenge is not so much to improve the precision in (1+1)d than to go to higher dimensions. It seems natural to tackle $\phi^4_{3}$ next, to increase difficulty progressively. Indeed, it is still a well defined super-renormalizable QFT, although its construction is slightly less trivial. The IR critical point is believed to also belong to the Ising universality class, and thus its critical exponents, which are useful for sanity checks, are known to very good precision thanks to impressive progress in the conformal boostrap program \cite{elshowk2012,elshowk2014,Kos:2016ysd, poland2019}.

Increasing the number of dimensions does not present any major conceptual difficulties for the lattice Monte-Carlo approach, but all non-perturbative deterministic approaches suffer in one way or another. We compare their prospects with those of our own approach.

Renormalized Hamiltonian truncation methods are promising but encounter a number of difficulties. First, in (2+1)d, the $\lambda\phi^4$ coupling term is less relevant, which makes the method \emph{a priori} converge slower. Second, it is no longer the case that the continuum Hamiltonian can be defined straightforwardly with normal ordering, and additional counter terms are needed in (2+1)d, requiring a non-trivial regularization (see however the recent advance \cite{miro2020}). Finally, the number of momenta to consider for a fixed energy cutoff is squared from (1+1)d to (2+1)d.

Matrix product state are easily generalized to projected entangled pair states (PEPS) in (2+1)d. The latter have been used to solve non-trivial problems to impressive precision \cite{vanderstraeten2016,liao2019,corboz2018,rader2018} but on the lattice and sufficiently away from criticality. The method is variational and limited to low bond dimensions beyond which the convergence is too slow. Near the continuum limit, the correlation length becomes infinite and we expect that larger physical and bond dimensions will be required, making PEPS less accurate. 

Generalizing our method to higher dimensions is also not straightforward. In (2+1)d, this would amount to contracting efficiently a cubic tensor network. To do so, we would proceed as in (1+1)d, by combining local truncations that remove short-range correlations with a coarse-graining procedure.  Although the \textsf{Gilt} algorithm is also applicable for three-dimensional networks,\footnote{As a matter of fact, the \textsf{Gilt} algorithm was precisely designed with in mind the goal of being amenable to higher dimensions (see \cite{hauru2018} for more details).} the corresponding numerical cost grows stiffly from $\mathcal{O}(\chi^7)$ to $\mathcal{O}(\chi^{12})$. Indeed, the $\textsf{Gilt}$ matrices must now be computed with respect to a cubic environment so as to remove three-dimensional short-range correlations.   
Due to this rather challenging cost, the performance of such an algorithm has not been conclusively demonstrated yet. 

In the end, for strongly coupled QFT in (2+1)d and \emph{a fortiori} (3+1)d, no deterministic method has a demonstrable edge so far. However, we believe that tensor network renormalization approaches, which are accurate for universal and non universal quantities in (1+1)d and suffer from no crippling difficulty in (2+1)d, are a promising option that deserves to be explored further.

\bigskip \medskip
\noindent
\emph{CD would like to thank Markus Hauru for countless discussions regarding the renormalization of tensor networks, collaboration on closely related topics, as well as useful comments about this manuscript. This project has received funding from the European Research Council (ERC) under the European Union’s Horizon 2020 research and innovation programme through the ERC Starting Grant WASCOSYS (No. 636201), as well as the Deutsche Forschungsgemeinschaft (DFG, German Research Foundation) under Germany’s Excellence Strategy – EXC-2111 – 390814868.}

\titleformat{name=\section}
{\normalfont}
{\centering {\textsc{App.} \thesection \; \raisebox{1pt}{\textbar} \;}}
{0pt}
{\normalsize\bfseries\centering}
\newpage
\appendix
\section{More on the field discretization}\label{appendix:discretization}

\noindent
Let us bound the coefficients of $T_{abcd}$. First, because $\log(n!)$ is convex, we have that
\begin{equation}
    \frac{1}{\sqrt{a!\, b!\, c!\, d!}} \leq \frac{1}{\left[(n/2)!\right]^2}\; ,
\end{equation}
where $n=(a+b+c+d)/2$, so that $T$ is dominated by its diagonal.
Next, so long as $\mu_0^2> -4$ (which is always verified near the continuum limit where $\mu_0\rightarrow 0)$, $h$ is easily bounded:
\begin{align*}
    h(n,\mu_0,\lambda) 
    &:=
    \int_\mathbb{R} \! \upd \phi\,\phi^{2n}
    \exp \!\bigg[ \Big( \! -\frac{\mu_0^2}{2}- 2 \Big) \, \phi^2  - \frac{\lambda}{4}\phi^4 \bigg] \\
    &\leq \;
    \int_\mathbb{R}\! \upd \phi\,\phi^{2n}\exp \! \bigg[- \frac{\lambda}{4}\phi^4 \bigg]\\
    &= \;\frac{1}{2} \Big(\frac{4}{\lambda}\Big)^{\frac{1}{4} + \frac{n}{2}} \, \underset{\leq \; (n/2)!}{\underbrace{\Gamma\Big[ \frac{1}{4} + \frac{n}{2}\Big]}}\, .
\end{align*}
Hence 
\begin{align}\label{eq:boundT}
    |T_{abcd}| \, \leq \,  \frac{1}{2} \Big(\frac{4}{\lambda}\Big)^{\frac{1}{4} + \frac{n}{2}} \frac{1}{(n/2)!} \, ,
\end{align}
which, \eg from Stirling's formula, goes to zero faster than exponentially.

This decay is only asymptotic and for $\lambda$ small enough, the coefficients of $T$ first grow exponentially, before they are killed off by the factorial. The smaller $\lambda$, that is the closer to the continuum limit we are, the later this tipping point occurs. This crossover can be estimated crudely with Stirling's formula:
\begin{equation}
    \log \! \bigg[\Big(\frac{4}{\lambda}\Big)^{ \frac{1}{4} + \frac{n}{2}} \frac{1}{(n/2)!}\bigg] \, \sim \,  \frac{n}{2} \log\Big(\frac{8}{\lambda n}\Big) \,,
\end{equation}
which is negative when $n\geq 8/\lambda$.
Thus, we see that $K_\text{max}$ has to grow only linearly in $\lambda^{-1}$ to insure that we are in the right asymptotic regime where the truncation error decreases super-exponentially.

As we go to the continuum limit and $\lambda\rightarrow 0$, we can get an estimate of the relevant range $\llbracket K_\text{min},K_\text{max}\rrbracket$ of indices to consider. It is much narrower than the naive $\llbracket 0,K_\text{max}\rrbracket$, because $K_\text{min}$ also grows linearly with $\lambda^{-1}$ for a fixed error threshold. Indeed, up to a normalization and for even integers, the right-hand side of \eqref{eq:boundT} is the Poisson distribution of parameter $4/\lambda$. The latter converges to a Gaussian with standard deviation $\sigma\propto\lambda^{-1/2}$ when $\lambda\rightarrow 0$. Hence, as we go to the continuum limit and for a fixed error threshold,  $K=K_\text{max}-K_\text{min}$ grows at most as $\lambda^{-1/2}$. Equivalently, the bond dimension of the initial tensor grows linearly with the inverse lattice spacing $\mathsf a^{-1}$.

\section{Convergence in bond dimension \texorpdfstring{$\chi$}{chi} and choice of  \texorpdfstring{$\epsilon_\mathsf{Gilt}$}{epsilon Gilt}\label{sec:app_chiAndEps}}

\noindent
As mentioned in the main text, the error bars quoted in fig.~\ref{fig:fc} are obtained from the fluctuations of our results as a function of the bond dimension. As we increase the bond dimension, we numerically observe that our results converge with apparent exponentially decaying oscillations around a limiting value. Since we have no theoretical means to estimate the error generated throughout the simulation, we use the amplitude of these oscillations as a ballpark estimate, which is most likely an overestimation of the error induced by the algorithm when using the highest bond dimension, namely $\chi=110$. More precisely, we compute the critical coupling for $\chi = 90,95,100,105,110$ and use the standard deviation of these five points as error bar, which we center around the value at $\chi=110$.

It turns out that the explanation provided above is slightly more subtle in practice, due to the fact that the bond dimension is not the only parameter of our algorithm. Indeed, the \textsf{Gilt} component of our renormalization scheme is tuned via the threshold parameter $\epsilon_\textsf{Gilt}$, which we recall is responsible for the trade-off between global accuracy and local truncation. Although there are no rigorous theoretical arguments to decide which value this parameter should be set to, we have several concordant heuristics. 

From the definition \eqref{eq:GiltMatrix} of the \textsf{Gilt} matrix, we understand that if $\epsilon_\textsf{Gilt}$ is too small, then \textsf{Gilt} will not truncate sufficiently and UV information will creep into the IR, and if it is too large, then the local error will be too important. By simply monitoring the amount of truncation performed by \textsf{Gilt}, we can already determine that choosing $\epsilon_\textsf{Gilt}$ in the range $10^{-6}$ to $10^{-8}$ for bond dimensions in the range $[90,110]$ is suitable. Furthermore, $\epsilon_\textsf{Gilt}$ should be chosen such that the error caused by the algorithm remains dominated by the coarse-graining step. Otherwise, increasing the bond dimension would have a very limited effect. In practice, this signifies that there is no optimal choice of $\epsilon_\textsf{Gilt}$ given a wide range of $\chi$, and that as $\chi$ is increased, $\epsilon_\textsf{Gilt}$ should be decreased accordingly. 

Our last heuristic is the expectation that, given a fixed bond dimension, small variations around the optimal value of $\epsilon_\textsf{Gilt}$ should have a negligible effect on the renormalization flow. In other words, we expect to find a range of values for $\epsilon_\textsf{Gilt}$ that yield the same final result. Naturally, the span of this range depends on the bond dimension we use. We do observe this phenomenon in practice and further notice that, close to the optimal value, a smaller $\epsilon_\textsf{Gilt}$ tends to favour the symmetry-broken phase, whereas a larger $\epsilon_\textsf{Gilt}$ tends to favour disordered phase. As such, the optimal $\epsilon_\textsf{Gilt}$ is typically found at a stationary point.

Concretely, we applied this program to determine the optimal $\epsilon_\textsf{Gilt}$ for bond dimensions $\chi=90,95,100,105,110$. In order to simplify the presentation, we made the choice of using the same $\epsilon_\textsf{Gilt}(\chi)$ for every $\lambda$.  Although this might not be optimal for every $\lambda$, we ensured that these values perform well for the whole range we consider. We found that $\epsilon_\textsf{Gilt}(90)=10\cdot 10^{-8}$, $\epsilon_\textsf{Gilt}(95)=8.5 \cdot 10^{-8}$ $\epsilon_\textsf{Gilt}(100)=7 \cdot 10^{-8}$, $\epsilon_\textsf{Gilt}(105)=6 \cdot 10^{-8}$ and $\epsilon_\textsf{Gilt}(110)=5 \cdot 10^{-8}$ fulfill all our requirements.

Finally, let us point out that this whole procedure is very specific to the algorithm we use. Indeed, by choosing $\epsilon_\textsf{Gilt}$, we only decide indirectly the bond dimension to which the bonds are truncated to during the \textsf{Gilt} step. Although this requires the preliminary study presented in this appendix, this approach has the advantage of being particularly versatile. Indeed, the bond dimension adapts automatically according to whether truncation can be done with more or less error, which depends on the amount of short-range information available. This turns out to be particularly relevant for our study as the initial tensors for the smallest values of $\lambda$ typically have a bond dimension that is larger than the maximal one that can be numerically dealt with during the coarse-graining step. We observe that the first step of \textsf{Gilt} significantly reduces the bond dimension and allows the algorithm to proceed.

\bigskip \medskip
\bibliography{main}

\end{document}